\documentclass{acmtrans2e}
\usepackage{epsf}
\usepackage{pst-rel-points}
\usepackage{colortab}
\usepackage{colordvi}
\usepackage{ifthen}
\usepackage{longtable}
\usepackage[normalem]{ulem}
\usepackage{amsmath}
\usepackage{amssymb}
\usepackage{amsfonts}
\usepackage{mathptm}
\usepackage{graphpap}
\usepackage{color,pstcol}
\usepackage{psboxit}
\usepackage{rotating}
\usepackage{pstcol}
\usepackage{pst-grad}
\usepackage{pstricks}
\usepackage{pst-text}
\usepackage{pst-node}
\usepackage{pst-coil}
\usepackage{epsfig}
\usepackage{url}

\newcommand{\cmt}[1]{}

\newtheorem{lemma}{Lemma}[section]
\newtheorem{theorem}{Theorem}[section]
\newdef{definition}[theorem]{Definition}
\newdef{corollary}[theorem]{Corollary}
\newcommand{\mybox}{\hfill$\Box$}
\newtheorem{exmpl}{Example}[section]
\newenvironment{example}{\begin{exmpl}\em}{\end{exmpl}}

\protect{}
\protect{}
\protect{\newlength{\BRACE}}
\protect{\newlength{\TAL}} 
\protect{\newlength{\BRACEL}}
\protect{\newcounter{programline}}

\newcommand{\PROGRAMBGN}{\setcounter{programline}{0}\begin{tabular}{|r@{.\ }l|}%
                        \hline 
                        }

\newcommand{\PROGRAMEND}{\rule[-8pt]{0pt}{12pt}\\ \hline \end{tabular}}

\newcommand{\entrynode}{\mbox{\sf\em Entry\/}}
\newcommand{\exitnode}{\mbox{\sf\em Exit\/}}
\newcommand{\Global}{\mbox{\sf\em Globals\/}}
\newcommand{\param}{\mbox{\sf\em Params\/}}

\newcommand{\lptr}{\mbox{\em lptr\/}}
\newcommand{\rptr}{\mbox{\em rptr\/}}

\newcommand{\myarrow}{\mbox{%
 \psset{unit=1cm}%
 \begin{pspicture}(0,0)(.27,.2)%
 \psline[linewidth=.15mm](0,.1)(.125,.1)(.125,.035)(.25,.1)(.125,.165)(.125,.1)%
 \end{pspicture}}%
}

\newcommand{\extend}[2]{\mbox{$#1\#\, #2$}}
\newcommand{\minus}{\ominus}
\newcommand{\Base}{\mbox{\sf\em Base\/}}
\newcommand{\Front}{\mbox{\sf\em Frontier\/}}
\newcommand{\appendG}[2]{\mbox{$#1 \cdot #2$}}

\newcommand{\cupG}{{\;\uplus\;}}
\newcommand{\bigcupG}{{\biguplus}}
\newcommand{\subG}[2]{\mbox{{\em RG}($#1$, $#2$)}}

\newcommand{\graphA}[1]{\mbox{{\sf\em\small GAll\/}$({#1})$}}
\newcommand{\graphO}[1]{\mbox{{\sf\em\small GOnly\/}$({#1})$}}
\newcommand{\newGraphO}[1]{\mbox{{\sf\em\small GOnly\/}$({#1})$}}

\newcommand{\states}[2]{\mbox{{\sf\em CFN\/}$({#1},{#2})$}}
\newcommand{\cNodes}[3]{\mbox{{\sf\em CN\/}$({#1},{#2},{#3})$}}
\newcommand{\clean}{\mbox{\sf\em CleanUp\/}}

\newcommand{\LnG}{\mbox{\sf LnG}}
\newcommand{\LnA}{\mbox{\sf LnA}}

\newcommand{\AP}[1]{\mbox{\sf\em P$({#1})$}}
\newcommand{\Target}{\mbox{\sf\em Target\/}}
\newcommand{\NULL}{\mbox{\sf\em null\/}}
\newcommand{\new}{\mbox{\sf\em new\/}}
\newcommand{\use}{\mbox{\sf\em use\/}}
\newcommand{\return}{\mbox{\sf\em return\/}}
\newcommand{\Empty}{\mbox{$\mathcal{E}$}}
\newcommand{\EFG}{\mbox{$\scalebox{1.2}{$\epsilon$}_{RG}$}}

\newcommand{\In}[1]{\mbox{\sf In$_#1$}}
\newcommand{\Out}[1]{\mbox{\sf Out$_#1$}}

\newcommand{\Aset}{\mbox{\sf\em AS\/}}
\newcommand{\uniquepath}{\mbox{\sf\em UniqueAccessPath$\!$?\/}}
\newcommand{\RootVar}{\mbox{\sf\em Root\/}}
\newcommand{\FieldName}{\mbox{\sf\em Field\/}}
\newcommand{\lNode}[1]{\mbox{{\sf\em lastNode\/}$({#1})$}}

\newcommand{\cannullify}{\mbox{\em Nullable}}
\newcommand{\nullify}{\mbox{\em Nullify}}
\newcommand{\cand}{\mbox{\em Candidates}}
\newcommand{\outfield}{\mbox{\em OutField}}

\newcommand{\Transp}{\mbox{\sf\em Transp\/}}
\newcommand{\Lhs}{\mbox{\sf\em lhs\/}}

\newcommand{\Lin}[1]{\mbox{{\sf E$\mathbb{L}$In}$_{#1}$}}
\newcommand{\Lout}[1]{\mbox{{\sf E$\mathbb{L}$Out}$_{#1}$}}
\newcommand{\Lgen}[1]{\mbox{{\sf E$\mathbb{L}$Gen}$_{#1}$}}
\newcommand{\Lkill}[1]{\mbox{{\sf E$\mathbb{L}$KillPath}$_{#1}$}}
\newcommand{\LD}[1]{\mbox{\sf\em LDirect$_{#1}$}}
\newcommand{\LT}[1]{\mbox{\sf\em LTransfer$_{#1}$}}

\newcommand{\CLin}[1]{\mbox{{\sf $\mathbb{L}$In}$_{#1}$}}
\newcommand{\CLout}[1]{\mbox{{\sf $\mathbb{L}$Out}$_{#1}$}}

\newcommand{\TLin}[1]{\mbox{{\sf T$\mathbb{L}$In}$_{#1}$}}

\newcommand{\NodeAin}{\mbox{{\sf $\mathbb{A}$In}}}

\newcommand{\NodeAout}{\mbox{{\sf $\mathbb{A}$Out}}}

\newcommand{\Avin}[1]{\mbox{{\sf $\mathbb{A}$vIn}$_{#1}$}}
\newcommand{\Avout}[1]{\mbox{{\sf $\mathbb{A}$vOut}$_{#1}$}}

\newcommand{\AVK}{\mbox{$Av{\mbox{\em Kill}}$}}
\newcommand{\AVD}{\mbox{$Av{\mbox{\em Direct}}$}}
\newcommand{\AVT}{\mbox{$Av{\mbox{\em Transfer}}$}}

\newcommand{\Antin}[1]{\mbox{{\sf $\mathbb{A}$nIn}$_{#1}$}}
\newcommand{\Antout}[1]{\mbox{{\sf $\mathbb{A}$nOut}$_{#1}$}}

\newcommand{\ANTK}{\mbox{$An{\mbox{\em Kill}}$}}
\newcommand{\ANTD}{\mbox{$An{\mbox{\em Direct}}$}}
\newcommand{\ANTT}{\mbox{$An{\mbox{\em Transfer}}$}}

\newcommand{\ELPK}{\mbox{$L{\mbox{\em Kill}}$}}
\newcommand{\ELPD}{\mbox{$L{\mbox{\em Direct}}$}}
\newcommand{\ELPT}{\mbox{$L{\mbox{\em Transfer}}$}}

\newcommand{\Pa}{\mbox{$p_{a}$}}
\newcommand{\Pb}{\mbox{$p_{b}$}}
\newcommand{\Porg}{\mbox{$p^o$}}
\newcommand{\Pmod}{\mbox{$p^m$}}
\newcommand{\Paorg}{\mbox{$p^o_a$}}
\newcommand{\Pborg}{\mbox{$p^o_b$}}
\newcommand{\Pamod}{\mbox{$p^m_a$}}
\newcommand{\Pbmod}{\mbox{$p^m_b$}}
\newcommand{\Slive}{\mbox{\em StatementLiveness\/}}
\newcommand{\Plive}{\mbox{\em PathLiveness\/}}
\newcommand{\live}{\mbox{\em Liveness\/}}
\newcommand{\Summary}{\mbox{\em Summary\/}}
\newcommand{\prefix}{\mbox{\em Prefixes\/}}

\newcommand{\Savail}{\mbox{\em StatementAvail\/}}
\newcommand{\Pavail}{\mbox{\em PathAvail\/}}
\newcommand{\avail}{\mbox{\em Avail\/}}

\newcommand{\Sant}{\mbox{\em StatementAnt\/}}
\newcommand{\Pant}{\mbox{\em PathAnt\/}}
\newcommand{\ant}{\mbox{\em Ant\/}}

\title{Heap Reference Analysis Using Access Graphs} 
\author{UDAY P. KHEDKER, 
	AMITABHA SANYAL and 
	AMEY KARKARE \\
	Department of Computer Science \& Engg., IIT Bombay.}

\begin{abstract}
Despite significant progress in the theory and practice of program
analysis, analyzing properties of heap data has not reached the same
level of maturity as the analysis of static and stack data.  The
spatial and temporal structure of stack and static data is well
understood while that of heap data seems arbitrary and is unbounded.
We devise bounded representations which summarize properties of the
heap data. This summarization is based on the structure of the program
which manipulates the heap.  The resulting summary representations are
certain kinds of graphs called {\em access graphs}. The boundedness of
these representations and the monotonicity of the operations to
manipulate them make it possible to compute them through data flow
analysis.

An important  application which benefits from  heap reference analysis
is  garbage  collection, where  currently  liveness is  conservatively
approximated by reachability from program variables. As a consequence,
current garbage collectors leave a  lot of garbage uncollected, a fact
which has been confirmed by several empirical studies.  We propose the
first ever end-to-end static analysis to distinguish live objects from
reachable  objects.  We  use  this information  to  make dead  objects
unreachable by modifying the  program. This application is interesting
because  it requires  discovering data  flow  information representing
complex  semantics.  In particular,  we  formulate  the following  new
analyses for heap data: liveness, availability, and anticipability and
propose  solution  methods for  them.   Together,  they cover  various
combinations of directions of analysis (i.e. forward and backward) and
confluence of information (i.e. union and intersection).
Our analysis can also be used for plugging memory leaks in 
C/C++ languages.
\end{abstract}

\boldmath

\begin{document}
\begin{bottomstuff}
\end{bottomstuff}

\maketitle

\section{Introduction}
\label{sec:intro}

Conceptually, data in a program is allocated in either the static data
area, stack, or heap.  Despite significant progress in the theory and
practice of program analysis, analyzing the properties of heap data
has not reached the same level of maturity as the analysis of static
and stack data.  Section~\ref{sec:back} investigates possible reasons.

In order to facilitate a systematic analysis, we devise bounded
representations which summarize properties of the heap data. This
summarization is based on the structure of the program which
manipulates the heap.  The resulting summary representations are
certain kinds of graphs, called access graphs which are obtained
through data flow analysis. We believe that our technique of
summarization is general enough to be also used in contexts other than
heap reference analysis.

\subsection{Improving Garbage Collection through Heap Reference Analysis}

An important application which benefits from heap reference analysis
is garbage collection, where liveness of heap data is conservatively
approximated by reachability. This amounts to approximating the future
of an execution with its past.  Since current garbage collectors
cannot distinguish live data from data that is reachable but not live,
they leave a lot of garbage uncollected.  This has been confirmed by
empirical
studies~\cite{hirz02,Hirzel.liveness.02,shah00,shah01,shah02} which
show that a large number (24\% to 76\%) of heap objects  which are
reachable at a program point are actually not accessed beyond that
point.  In order to collect such objects, we perform static analyses
to make dead objects unreachable by setting appropriate references to
\NULL. The idea that doing so would facilitate better garbage
collection is well known as ``Cedar Mesa Folk Wisdom''~\cite{gcfaq}.
The empirical attempts at achieving this have
been~\cite{shah01,shah02}.

Garbage collection is an interesting application for us because it
requires discovering data flow information representing complex
semantics.  In particular, we need to discover four properties of heap
references: liveness, aliasing, availability, and anticipability.
Liveness captures references that may be used beyond the program point
under consideration. Only the references that are not live can be
considered for \NULL\ assignments.  Safety of \NULL\ assignments
further requires (a) discovering all possible ways of accessing a
given heap memory cell (aliasing), and (b) ensuring that the reference
being nullified is accessible (availability and anticipability).

\begin{figure}[t]
\rule{\textwidth}{.2mm}
\begin{tabular}{@{}c@{}|@{}c}
\begin{tabular}{@{}c@{}@{}}
{
$\begin{array}{ll}
\\
    1. & w = x 
    {\parbox{1.75in}{\raggedleft  // $x$ points to $m_a$ }}\\
    2. & \mbox{\em while } (x.getdata() < max) \\
       & \{ \\
    3. & \;\;\;\;\;\;\;\; x = x.\rptr  \\
       & \} \\
    4. & y = x.\lptr  \\ 
    5. & z = \mbox{\em New } \mbox{\em class\_of\_z}
    \parbox{1.4in}{\raggedleft  // Possible GC Point} \\
    6. & y = y.\lptr  \\
    7. & z.sum = x.{\lptr}.getdata() + y.getdata()   \\ \\
  \end{array}$
}
\\
{(a) A Program Fragment}
\\ \hline
\multicolumn{1}{@{}l@{}}{
\begin{pspicture}(-6.6,-6)(-.5,-.9)
\psset{xunit=.6cm}
\psset{yunit=.6cm}
{
\psline[linestyle=dashed](-9.75,-7.5)(-9.75,-1.75)
\psline(-9.25,-7.2)(-7,-7.2) \rput(-6.25,-7.2){Heap} \psline{->}(-5.5,-7.2)(-3,-7.2)
\rput(-10.5,-7.2){Stack} 
\rput(-10.5,-2.5){\rnode{n0}{\psframebox{$z$}}}
\rput(-10.5,-3.75){\rnode{n0}{\psframebox{$x$}}}
\rput(-10.5,-4.75){\rnode{w}{\psframebox[framesep=.09]{$w$}}}
\rput(-9,-4.5){\rnode{n1}{\pscirclebox[fillcolor=lightgray,framesep=0]{$m_a$}}}
\ncline{->}{w}{n1}
\rput(-10.5,-6){\rnode{n00}{\psframebox{$y$}}}
\rput(-9,-2.5){\rnode{n11}{\pscirclebox[framesep=0]{$m_k$}}}
\rput(-7.5,-3.75){\rnode{n2}{\pscirclebox[fillcolor=lightgray,framesep=0]{$m_b$}}}
\nccurve[angleA=0,angleB=170,linewidth=.6mm,linestyle=dashed,dash=1mm .6mm]{->}{n0}{n2}
\nccurve[angleA=0,angleB=150,linewidth=.6mm,linestyle=dashed,
	dash=1mm .5mm]{->}{n0}{n1}
\ncline{->}{n1}{n2}
\aput[0]{:U}{$\rptr$}
\rput(-6,-3){\rnode{n3}{\pscirclebox[framesep=0]{$m_c$}}}
\ncline{->}{n2}{n3}
\aput[0]{:U}{$\rptr$}
\nccurve[angleA=2,angleB=170,linewidth=.6mm,linestyle=dashed,
	dash=1mm .6mm]{->}{n0}{n3}
\rput(-4.5,-2.25){\rnode{n4}{\pscirclebox[framesep=0]{$m_d$}}}
\ncline{->}{n3}{n4}
\aput[0]{:U}{$\rptr$}
\nccurve[angleA=10,angleB=170,linewidth=.6mm,linestyle=dashed,
	dash=1mm .6mm]{->}{n0}{n4}
\rput(-3,-3){\rnode{n5}{\pscirclebox[framesep=0]{$m_e$}}}
\ncline[linewidth=.7mm]{->}{n4}{n5}
\aput[0]{:U}{$\lptr$}
\rput(-6,-4.5){\rnode{n6}{\pscirclebox[fillcolor=lightgray,framesep=0]{$m_f$}}}
\ncline[linewidth=.7mm]{->}{n2}{n6}
\aput[0]{:U}{$\lptr$}
\rput(-4.5,-3.75){\rnode{n7}{\pscirclebox[framesep=0]{$m_g$}}}
\ncline[linewidth=.7mm]{->}{n3}{n7}
\aput[0]{:U}{$\lptr$}
\rput(-4.5,-5.15){\rnode{n8}{\pscirclebox[fillcolor=lightgray,framesep=0]{$m_h$}}}
\ncline[linewidth=.7mm]{->}{n6}{n8}
\aput[0]{:U}{$\lptr$}
\rput(-7.5,-5.25){\rnode{n9}{\pscirclebox[fillcolor=lightgray,framesep=0]{$m_i$}}}
\ncline[linewidth=.7mm]{->}{n1}{n9}
\aput[0]{:U}{$\lptr$}
\rput(-5.75,-6){\rnode{n10}{\pscirclebox[framesep=0]{$m_j$}}}
\ncline[linewidth=.7mm]{->}{n9}{n10}
\aput[0]{:U}(.3){$\lptr$}
\nccurve[angleA=0,angleB=240,linewidth=.6mm,linestyle=dashed,dash=1mm .6mm,nodesepB=-.08]{->}{n00}{n6}
\nccurve[angleA=0,angleB=200,linewidth=.6mm,linestyle=dashed,dash=1mm .6mm]{->}{n00}{n9}
\nccurve[angleA=-10,angleB=240,linewidth=.6mm,linestyle=dashed,dash=1mm .6mm,nodesepB=-.08]{->}{n00}{n7}
\nccurve[angleA=-20,angleB=250,linewidth=.6mm,linestyle=dashed,dash=1mm .6mm,ncurv=.95]{->}{n00}{n5}
\rput(-2.25,-2.65){\rnode{n51}{}}
\ncline{->}{n5}{n51}
\rput(-1.65,-3.75){\rnode{n51}{\pscirclebox[framesep=0]{$m_l$}}}
\ncline[linewidth=.7mm]{->}{n5}{n51}
\aput[0]{:U}(.5){$\lptr$} 

\rput(-2.75,-4.5){\rnode{n51}{\pscirclebox[framesep=0]{$m_m$}}}
\ncline[linewidth=.7mm]{->}{n7}{n51}
\aput[0]{:U}(.5){$\lptr$} 

\rput(-3.75,-3.4){\rnode{n51}{}}
\ncline{->}{n7}{n51}
\rput(-3.75,-5.6){\rnode{n51}{}}
\ncline{->}{n8}{n51}
\rput(-3.75,-4.9){\rnode{n51}{}}
\ncline{->}{n8}{n51}
\rput(-4.85,-6.35){\rnode{n51}{}}
\ncline{->}{n10}{n51}
\rput(-4.85,-5.65){\rnode{n51}{}}
\ncline{->}{n10}{n51}
\rput(-8.25,-2.85){\rnode{n51}{}}
\ncline{->}{n11}{n51}
\rput(-8.25,-2.15){\rnode{n51}{}}
\ncline{->}{n11}{n51}
\rput[l](-11.,-9.){(b) \parbox[t]{2.2in}{
Superimposition of memory graphs before line 5.  Dashed arrows capture
the effect of different iterations of the {\em while} loop.  All thick
arrows (both dashed and solid) are live links.}}  }
\end{pspicture}
}
\end{tabular}
&
\renewcommand{\arraystretch}{1.1}
\begin{tabular}{@{}ll@{}}
\\
\LCC & \lightgray\\
   & {$y = z = \NULL$}\\
\ECC 
1. & $w = x $\\
\LCC & \lightgray\\
 & {$w = \NULL$}\\
\ECC 
2. & {\em while } $(x.getdata() < max)$\\
\LCC & \lightgray\\
   & \{$\;\;\;\;\;\; 
      x.\lptr = \NULL$\\
\ECC
3. & $\;\;\;\;\;\;\;\;\, x = x.\rptr$\\
& \} \\
\LCC & \lightgray\\
  & {$x.\rptr = x.\lptr.\rptr = \NULL$}\\
   & {$x.\lptr.\lptr.\lptr = \NULL$}\\
   & {$x.\lptr.\lptr.\rptr = \NULL$}\\
\ECC
4. & $y = x.\lptr$\\ 
\LCC & \lightgray\\
   & {$y.\rptr = y.\lptr.\lptr = y.\lptr.\rptr = \NULL$}\\
\ECC
5. & $z$ = {\em New } {\em class\_of\_z}\\
\LCC & \lightgray\\
   & {$z.\lptr = z.\rptr = \NULL$}\\
\ECC
6. & $y = y.\lptr$\\
\LCC & \lightgray\\
   & $x.\lptr.\lptr = y.\lptr = y.\rptr = \NULL$\\
\ECC
7. & $z.sum = x.\lptr.getdata() + y.getdata()$\\
\LCC & \lightgray\\
   & {$x = y = z = \NULL$}\\
\ECC
\\
\multicolumn{2}{@{}c@{}}{(c) \parbox[t]{2.25in}{The
    modified program. Highlighted statements indicate the \NULL\
    assignments inserted in the program using our method. (More details
    in Section~\ref{sec:nullability})}}
\end{tabular}
\end{tabular}
\caption{A motivating example.}
\label{fig:memory.graph}
\rule{\textwidth}{.2mm}
\end{figure}
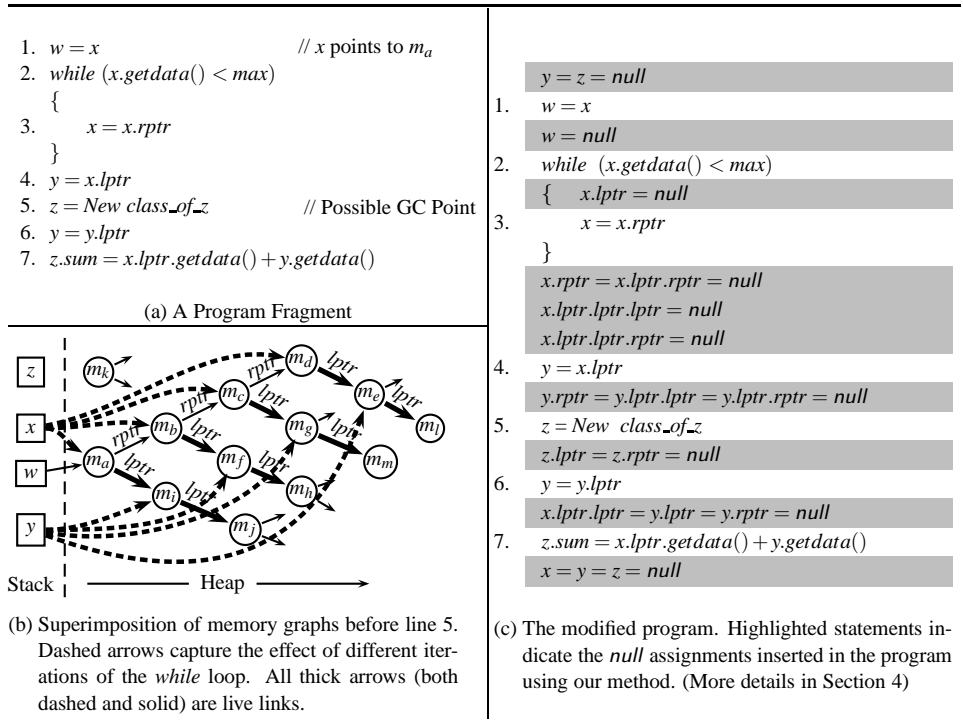

For simplicity of exposition, we present our method using a memory model
similar to that of Java.  {Extensions required for handling C/C++ model of
heap usage are easy and are explained in Section~\ref{sec:c++ext}.}
We assume that
root variable references are on the stack and the actual objects
corresponding to the root variables are in the heap.  In the rest of
the paper we ignore non-reference variables.  We view the heap at a
program point as a directed graph called {\em memory graph}. Root
variables form the entry nodes of a memory graph.  Other nodes in the
graph correspond to objects on the heap and edges correspond to
references. The out-edges of entry nodes are labeled by root variable
names while out-edges of other nodes are labeled by field names. The
edges in the memory graph are  called {\em links}.

\begin{example}
\label{exmp:motivation}
Figure~\ref{fig:memory.graph} shows a program fragment and its memory
graphs before line 5. Depending upon the number of times the {\em
while} loop is executed $x$ points to $m_a$, $m_b$, $m_c$ etc.
Correspondingly, $y$ points to $m_i$, $m_f$, $m_g$ etc.  The call to
{\em New\/} on line 5 may require garbage collection.  A conventional
copying collector will preserve all nodes except $m_{k}$. However,
only a few of them are used beyond line 5.

The modified program is an evidence of the strength of our approach.
It makes the unused nodes unreachable by nullifying relevant links.
The modifications in the program are general enough to nullify
appropriate links for any number of iterations of the loop.  Observe
that a \NULL\ assignment has also been inserted within the loop body
thereby making some memory unreachable in each iteration of the loop.
\mybox
\end{example}

After such modifications, a garbage collector will collect a lot more
garbage. Further, since copying collectors process only live data,
garbage collection by such collectors will be faster. Both these
facts are corroborated by our empirical measurements
(Section~\ref{sec:measurements}).

In the context of C/C++, instead of setting the references to \NULL,
allocated memory will have to be explicitly deallocated after checking
that no alias is live.

\subsection{Difficulties in Analyzing Heap Data}
\label{sec:back}

A program accesses data through expressions which have l-values and
hence are called {\em access expressions}. They can be scalar
variables such as $x$, or may involve an array access such as
$a[2*i]$, or can be a reference expression such as $x.l.r$.

An important question that any program analysis has to answer is: {\em
Can an access expression $\alpha_1$ at program point~$p_1$ have the
same l-value as $\alpha_2$ at program point~$p_2$?}  Note that the
access expressions {or} program points could be identical.  The
precision of the analysis depends on the precision of the answer to
the above question.

When the access expressions are simple and correspond to scalar data,
answering the above question is often easy because, the mapping of
access expressions to l-values remains fixed in a given scope
throughout the execution of a program.  However in the case of array
or reference expressions, the mapping between an access expression and
its l-value is likely to change during execution. From now on, we
shall limit our attention to reference expressions, since these are
the expressions that are primarily used to access the heap.  Observe that
manipulation of the heap is nothing but changing the mapping between
reference expressions and their l-values. For example, in
Figure~\ref{fig:memory.graph}, access expression $x.\lptr$ refers to
$m_i$ when the execution reaches line number 2 and may refer to $m_i$,
$m_f$, $m_g$, or $m_e$ at line 4.

This implies that, subject to type compatibility, any access
expression can correspond to any heap data, making it difficult to
answer the question mentioned above.  The problem is compounded
because the program may contain loops implying that the same access
expression appearing at the same program point may refer to different
l-values at different points of time.  Besides, the heap data may
contain cycles, causing an infinite number of access expressions to
refer to the same l-value.  All these make analysis of programs
involving heaps difficult.

\subsection{Contributions of This Paper}

The contributions of this paper fall in the following two categories
\begin{itemize}
\item {\em  Contributions in Data  Flow Analysis.}  We present  a data
flow framework in which the data flow values represent abstractions of
heap.   An interesting  aspect  of our  method  is the  way we  obtain
bounded representations  of the properties  by using the  structure of
the  program which  manipulates the  heap.  As  a consequence  of this
summarization,  the  values  of  data flow  information  constitute  a
complete  lattice  with  finite  height. Further,  we  have  carefully
identified a set of monotonic  operations to manipulate this data flow
information.  Hence, the standard results of data flow analysis can be
extended to  heap reference  analysis. Due to  the generality  of this
approach, it can be applied to other analyses as well.

\item {\em Contributions in Heap Data Analysis.}  We propose the first
ever end-to-end solution (in the intraprocedural context) for
statically discovering heap references which can be made \NULL\ to
improve garbage collection. The only approach which comes close to our
approach is the {\em heap safety automaton\/} based
approach~\cite{ran.shaham-sas03}.  However, our approach is superior
to their approach in terms of completeness, effectiveness, and
efficiency (details in Section~\ref{sec:ran.comparison}).
\end{itemize}

The concept which unifies the contributions is the
summarization of heap properties which uses the fact that {\em the heap
manipulations consist of repeating patterns which bear a close
resemblance to the program structure.}  Our approach to summarization
is more natural and more precise than other approaches because it does
not depend on an a-priori
bound~\cite{DBLP:conf/popl/JonesM79,jones82flexible,LarusH1988,ChaseWegZad90}.

\subsection{Organization of the paper}

The rest of the paper is organized as follows.
Section~\ref{sec:liveness} defines the concept of explicit liveness
of heap objects and formulates a data flow analysis
by using access graphs as data flow values.
Section~\ref{sec:other.analyses} defines other properties required for ensuring
safety of \NULL\ assignment insertion.
Section~\ref{sec:nullability} explains how \NULL\ assignments are
inserted.
Section~\ref{sec:termination} discusses convergence and complexity
issues.
Section~\ref{sec:soundness} shows the soundness of our approach.
Section~\ref{sec:measurements} presents empirical results.
Section~\ref{sec:c++ext} extends the approach to C++. 
Section~\ref{sec:related} reviews related work while
Section~\ref{sec:conclusions} concludes the paper.

\section{Explicit Liveness Analysis of Heap References}
\label{sec:liveness}

Our method discovers live links at each program point, i.e., links
which may be used in the program beyond the point under consideration.
Links which are not live can be set to \NULL. 
This section describes the liveness analysis. In particular, we define
liveness of heap references, devise a bounded representation called an 
{\em access graph\/} for liveness, and
then propose a data flow analysis for discovering liveness. 
Other analyses required for safety of \NULL\ insertion are described in 
Section~\ref{sec:other.analyses}.

Our method is flow sensitive but context insensitive. This means that 
we compute point-specific information in each procedure by taking into
account the flow of control at the intraprocedural level and 
by approximating the interprocedural information such that it is
not context-specific but is safe in all calling contexts.
For the purpose of analysis, arrays are handled by approximating any 
occurrence of an array element by the entire array.
The current version models exception handling by
explicating possible control flows. However, programs containing 
threads are not covered.

\subsection{Access Paths}

In order to discover
liveness and other properties of heap, we need a way of naming links in the
memory graph.  We do this using  access paths.

\label{sec:concepts.def}
An {\em access path\/} is a root variable name followed by a sequence
of zero or more field names and is denoted by \mbox{$\rho_x \equiv
x\myarrow f_1\myarrow f_2\myarrow\cdots\myarrow f_k$}.  Since an
access path represents a path in a memory graph, it can be used for
naming links and nodes.  An access path consisting of just a root
variable name is called a {\em simple\/} access path; it represents a
path consisting of a single link corresponding to the root
variable. \Empty\ denotes an empty access path.

The last field name in an access path $\rho$ is called its {\em
frontier\/} and is denoted by $\Front(\rho)$. The frontier of a simple
access path is the root variable name.  The access path corresponding
to the longest sequence of names in $\rho$ excluding its frontier is
called its {\em base\/} and is denoted by $\Base(\rho)$.  Base of a
simple access path is the empty access path \Empty. The object reached by traversing an
access path $\rho$ is called the {\em target\/} of the access path and
is denoted by $\Target(\rho)$.  When we use an access path $\rho$ to refer
to a link in a memory graph, it  denotes the last link in $\rho$, i.e. the
link corresponding to $\Front(\rho)$.
\begin{example}
\label{exmp:access.path.1}
As explained earlier, Figure~\ref{fig:memory.graph}(b) is the
superimposition of memory graphs that can result before line 5 for
different executions of the program.  For the access path \mbox{$\rho_x \equiv
x\myarrow\lptr\myarrow\lptr$}, depending on whether the {\em while\/}
loop is executed 0, 1, 2, or 3 times, $\Target(\rho_x)$ denotes nodes
$m_j$, $m_h$, $m_m$, or $m_l$.  $\Front(\rho_x)$ denotes one of the
links \mbox{$m_i\rightarrow m_j$}, \mbox{$m_f\rightarrow m_{h}$},
\mbox{$m_g\rightarrow m_{m}$} or \mbox{$m_e\rightarrow m_{l}$}.
$\Base(\rho_x)$ represents the following paths in the heap memory:
\mbox{$x\rightarrow m_a\rightarrow m_i$}, \mbox{$x\rightarrow
m_b\rightarrow m_f$}, \mbox{$x\rightarrow m_c\rightarrow m_g$} or
\mbox{$x\rightarrow m_d\rightarrow m_e$}. 
\mybox
\end{example}

In the rest of the paper, $\alpha$ denotes an access expression,
$\rho$ denotes an access path and $\sigma$ denotes a (possibly empty)
sequence of field names separated by $\myarrow$. Let the 
access expression 
$\alpha_x$ be $x.f_1.f_2\ldots f_n$. Then, the
corresponding access path $\rho_x$ is $x\myarrow f_1\myarrow f_2\ldots f_n$.
When the root variable name is not required,
we drop the subscripts from $\alpha_x$ and $\rho_x$.

\subsection{Program Flow Graph}

Since the current version of our method involves context insensitive analysis, 
each procedure is analyzed separately and only once. Thus there is no need of 
maintaining a call graph and we use the term program and procedure
interchangeably.

To simplify the description of analysis we make the following assumptions:
\begin{itemize}
\item The  program flow  graph has  a unique  $\entrynode$ and  a unique
      $\exitnode$ node. We assume that there is a distinguished {\tt main\/} procedure.
\item Each  statement  forms  a basic  block.
\item The conditions  that alter  flow of  control are  made up  only of
      simple variables.  If not,  the offending reference  expression is
      assigned to  a fresh simple  variable before the condition  and is
      replaced by the fresh variable in the condition.
\end{itemize}
With these simplification, each statement falls in one of the following
categories:
\begin{itemize}
\item {\em Function Calls\/}. These are statements 
	\mbox{$x = f(\alpha_y, \alpha_z, \ldots)$} where the functions 
      involve access expressions in arguments. The type of $x$ does not matter.
\item {\em Assignment Statements\/}.  These are assignments to
  references and are denoted by \mbox{$\alpha_x = \alpha_y$}.  Only
  these statements can modify the structure of the heap.
\item {\em Use Statements\/}. These statements use heap references to
  access heap data but do not modify heap references. For the purpose
  of analysis, these statements are abstracted as lists of expressions
  $\alpha_y.d$ where $\alpha_y$ is an access expression
  and $d$ is a non-reference. 
\item {\em Return Statement\/} of the type $\return \; \alpha_x$ involving
reference variable $x$.
\item {\em Other Statements\/}.  These statements include all
  statements which do not refer to the heap.  We ignore these
  statements since they do not influence heap reference analysis.
\end{itemize}

\newcommand{\onepath}{\psi}

When we talk  about the execution path, we shall  refer to the execution
of the program derived by  retaining all function calls, assignments and
use statements and ignoring the condition checks in the path.

For    simplicity   of    exposition,    we    present   the    analyses
assuming   that   there   are   no  cycles   in   the   heap.   
This assumption does not limit the theory in any way because 
our analyses inherently compute conservative information in the presence
of cycles without requiring any special treatment.

\subsection{Liveness of Access Paths}
\label{sec:liveness.specs}

A link  $l$ is {\em live\/}  at a program point  {$p$} if it is  used in
some control flow path starting from {$p$}. Note that $l$ may be used in
two different ways. It may be dereferenced to access an object or tested
for comparison. An  erroneous nullification of $l$ would  affect the two
uses in different  ways: Dereferencing $l$ would result  in an exception
being raised whereas testing $l$ for  comparison may alter the result of
condition and thereby the execution path.

Figure~\ref{fig:memory.graph}(b) shows links that are live before line 5
by thick arrows. For  a link $l$ to be live, there must  be at least one
access path from some root variable to  $l$ such that every link in this
path is  live. This is the  path that is actually  traversed while using
$l$.

Since our  technique involves nullification of  access paths, we  need to extend
the notion of liveness from links to  access paths. An access path is defined to
be  {\em  live\/}  {at  $p$}  if  the link  corresponding  to  its  frontier  is
 live  along  some  path  starting at  $p$.   Safety  of  \NULL\
assignments  requires that the  access paths  which are  live are  excluded from
nullification.

We initially  limit ourselves to a  subset of live access  paths, whose liveness
can be  determined without taking into  account the aliases  created before $p$.
These  access paths  are live  solely because  of the  execution of  the program
beyond  $p$.   We call  access  paths  which are  live  in  this  sense as  {\em
explicitly live} access paths. An interesting property of explicitly live access
paths is that they form the minimal set covering every live link. 

\begin{example}
\label{exmp:liveness}
If the body of the {\em while\/} loop in Figure~\ref{fig:memory.graph}(a) is not
executed   even  once,   \mbox{$\Target(y)=m_i$}  at   line  5   and   the  link
\mbox{$m_i\rightarrow m_j$} is live at line 5  because it is used in line 6. The
access paths \mbox{$y$} and \mbox{$y\myarrow \lptr$} are explicitly live because
their liveness at 5 can be determined solely from the statements from 5 onwards.
In contrast, the access path \mbox{$w\myarrow\lptr\myarrow\lptr$} is live without
being explicitly live.  It becomes  live because of the alias between \mbox{$y$}
and \mbox{$w\myarrow\lptr$} and this alias was created before 5.  Also note that
if an access path is explicitly live, so are all its prefixes.  \mybox
\end{example}

\begin{example}
\label{exmp:liveness.defn}
We illustrate  the issues  in determining explicit  liveness of access  paths by
considering the assignment \mbox{$x.r.n = y.n.n$}.


\begin{itemize}
\item {\em Killed Access Paths}. Since the assignment modifies $\Front(x\myarrow
r\myarrow  n)$,  any  access  path  which  is  live  after  the  assignment  and
 has $x\myarrow r\myarrow n$ as prefix will cease to be  live before the 
 assignment. Access paths that are live after  the assignment and not killed  
 by it are live  before the assignment also.
\item {\em Directly Generated Access Paths}. All prefixes of $x\myarrow r$ 
and $y\myarrow n$ are explicitly live before the assignment due to the local effect
of the assignment.
\item {\em Transferred Access Paths}. If 
\mbox{$x\myarrow r \myarrow n \myarrow \sigma$} is live after
  the assignment, then \mbox{$y\myarrow n \myarrow n \myarrow \sigma$}
  will be live before the assignment. For example, if \mbox{$x\myarrow
  r \myarrow n \myarrow n$} is live after the assignment, then
  \mbox{$y\myarrow n \myarrow n \myarrow n$} will be live before the
  assignment. The sequence of field names $\sigma$ is viewed as being
  {\em transferred\/} from {$x\myarrow r \myarrow n$} to
  \mbox{$y\myarrow n \myarrow n$}.
\mybox
\end{itemize}
\end{example}

We now define liveness by generalizing the above observations.
We use the notation \mbox{$\rho_x\myarrow *$} to enumerate all access paths
which have $\rho_x$ as a prefix. The summary liveness information for a set $S$ of
reference variables is defined as follows:
\begin{eqnarray*}
\Summary(S) & = & 
    \bigcup_{x \in S} \{ x\myarrow * \} 
\end{eqnarray*}
Further, the set of  all global variables is denoted by \Global\  and the set of
formal parameters of the function being analyzed is denoted by \param.

\begin{definition}{\rm\bf Explicit Liveness}.
\label{def:explicit.liveness}
The set of { explicitly} live access paths at a program point~$p$,
denoted by $\live_p$ is defined as follows.
  \begin{eqnarray*}
    \live_p& = & 
    \displaystyle\bigcup_{\onepath \in Paths(p)}(\Plive^{\onepath}_{p}) 
  \end{eqnarray*}
where, $\onepath \in Paths(p)$ is a control flow path $p$ to \exitnode\ and
$\Plive^{\onepath}_{p}$ denotes the
liveness at $p$ along $\onepath$ and is defined as follows. 
If $p$ is not program exit then let the statement which follows it be denoted by
  $s$ and the program point immediately following $s$ be denoted by $p'$. Then,
  \begin{eqnarray*}
    \Plive^{\onepath}_{p}& = & \left\{\begin{array}{cl}
    \emptyset & p \mbox{ is  \exitnode\ of {\tt main}} \\\
    \Summary(\Global) & p \mbox{ is  \exitnode\ of some procedure} \\
    \Slive_{s}(\Plive^{\onepath}_{p'})  & \mbox{otherwise}
    \end{array}\right. 
  \end{eqnarray*}
where the flow function for $s$ is defined as follows:
  \begin{eqnarray*}
    \Slive_{s}(X) & = & (X - \ELPK_s) \; \cup \; \ELPD_s \;\cup\;\ELPT_s(X) 
  \end{eqnarray*}
$\ELPK_s$ denotes the sets of access paths which cease to be live before 
statement $s$, $\ELPD_s$ denotes the set of access paths which become live due 
to local effect of $s$ and $\ELPT_s(X)$ denotes the
the set of access paths which become live before $s$ due to transfer of liveness from
live access paths after $s$. 
They are defined in Figure~\ref{fig:flow.fun.liveness}. \mybox
\end{definition}
Observe that
the    definitions   of   $\ELPK_s$,    $\ELPD_s$,   and
$\ELPT_s(X)$ ensure that the $\live_{p}$ is prefix-closed.

\begin{figure}[t]
\begin{center}
\scalebox{1.2}{%
$
\begin{array}{|l|c|c|c|}
\hline
\mbox{Statement } s & \ELPK_s & \ELPD_s & \ELPT_s(X) \\  \hline\hline
\alpha_x = \alpha_y & \{ \rho_x\myarrow * \} & 
	\prefix(\Base(\rho_x))\cup \prefix(
	\Base(\rho_y))  &
	\{ \rho_y\myarrow \sigma \mid \rho_x\myarrow \sigma \in X\} 
 		\\ \hline
 \alpha_x = f(\alpha_y) & \{\rho_x\myarrow *\} & 
 \prefix(\Base(\rho_x))  & \emptyset \\ 
& & \cup\; \Summary(\{\rho_y\}\cup\Global) & 
 		\\ \hline
\alpha_x = \new & \{ \rho_x\myarrow * \} & \prefix(\Base(\rho_x)) & \emptyset
 		\\ \hline
\alpha_x = \NULL & \{ \rho_x\myarrow * \} & \prefix(\Base(\rho_x)) & \emptyset
 		\\ \hline
\use  \; \alpha_y.d  & \emptyset & \prefix(\rho_y) & \emptyset
 		\\ \hline
\return \;  \alpha_y & \emptyset &  \Summary(\{\rho_y\}) & \emptyset 
 		\\ \hline
\mbox{other}  & \emptyset &  \emptyset & \emptyset 
 		\\ \hline
\end{array}
$}
\end{center}
\caption{Defining Flow Functions for Liveness. \Global\ denotes the set of global 
	references and \param\ denotes the set of formal parameters. 
	For simplicity, we have shown 
	a single access expression on the RHS.}
\label{fig:flow.fun.liveness}
\rule{\textwidth}{.2mm}
\end{figure}

\begin{example}
\label{exmp:unbounded.ap}
In Figure~\ref{fig:memory.graph},  it cannot be  statically determined
which  link is  represented by  access expression  \mbox{$x.\lptr$} at
line 4. Depending  upon the number of iterations  of the {\em while\/}
loop, it may be any of  the links represented by thick arrows. Thus at
line    1,   we    have   to    assume   that    all    access   paths
\mbox{\{\mbox{$x\myarrow\lptr\myarrow\lptr$},
\mbox{$x\myarrow\rptr\myarrow\lptr\myarrow\lptr$},
\mbox{$x\myarrow\rptr\myarrow\rptr\myarrow\lptr\myarrow\lptr$},
\ldots\}} are explicitly live.  \mybox
\end{example}

In general, an infinite number  of access paths with unbounded lengths
may be live before a  loop. Clearly, performing data flow analysis for
access    paths   requires    a   suitable    finite   representation.
Section~\ref{sec:access.graphs} defines access graphs for the purpose.

\subsection{Representing Sets of Access Paths by Access Graphs}
\label{sec:access.graphs}

In the  presence of loops, the set  of access paths may  be infinite and
the  lengths of access  paths may  be unbounded.   If the  algorithm for
analysis tries  to compute sets of access  paths explicitly, termination
cannot be  guaranteed.  We solve this  problem by representing  a set of
access paths by a graph of bounded size.

\subsubsection{Defining Access Graphs}
\label{sec:dependence}
An  {\em  access   graph},  denoted  by  $G_v$,  is   a  directed  graph
\mbox{$\langle n_0,  N, E \rangle$}  representing a set of  access paths
starting from a root variable $v$.\footnote{Where the root variable name
is not required, we drop the  subscript $v$ from $G_v$.}  $N$ is the set
of nodes, $n_0  \in N_F$ is the  entry node with no in-edges  and $E$ is
the set  of edges. Every  path in the  graph represents an  access path.
The {\em empty graph\/} $\Empty\!_G$ has  no nodes or edges and does not
accept any access path.

The entry node of an access graphs  is labeled with the name of the root
variable  while the  non-entry nodes  are  labeled with  a unique  label
created as  follows: If a  field name $f$  is referenced in  basic block
$b$,  we  create  an  access  graph node  with  a  label  \mbox{$\langle
f,b,i\rangle$} where $i$ is  the instance number used for distinguishing
multiple occurrences of the field name $f$ in block $b$.  Note that this
implies that  the nodes  with the same  label are treated  as identical.
Often, $i$ is 0 and in such a case we denote the label \mbox{$\langle f,
b, 0\rangle$} by $f_b$  for brevity.  Access paths \mbox{$\rho_x\myarrow
*$} are  represented by  including a summary  node denoted $n_*$  with a
self loop over it.  It is  distinct from all other nodes but matches the
field name of any other node.

A node  in the access graph represents  one or more links  in the memory
graph.  Additionally,  during analysis, it represents a  state of access
graph  construction (explained in  Section~\ref{sec:summarisation}).  An
edge \mbox{$  f_n\rightarrow g_m$} in  an access graph at  program point
$p$ indicates  that a  link corresponding to  field $f$  dereferenced in
block $n$ may  be used to dereference a link  corresponding to field $g$
in  block $m$  on  some path  starting at  $p$.  This has  been used  in
Section~\ref{sec:complexity} to argue that  the size of access graphs in
practical programs is small.

Pictorially, the entry node of an access graph is indicated by an
incoming double arrow.

\begin{figure}[t]
\small
\begin{pspicture}(2.,1.35)(10,4.15)
\psset{unit=1mm}
\psrelpoint{origin}{base}{0}{0}
\psrelpoint{base}{n0}{30}{25}
\rput(\x{n0},\y{n0}){\rnode{n0}{}}
\psrelpoint{n0}{n1}{0}{-6}
\psrelpoint{n1}{n2}{0}{-9}
\rput(\x{n1},\y{n1}){\rnode{n1}{1 \psframebox{$x = x.r$} \white 1}}
\rput(\x{n2},\y{n2}){\rnode{n2}{2 \psframebox{$x = x.r$} \white 2}}
\psrelpoint{n2}{n3}{0}{-6}
\rput(\x{n3},\y{n3}){\rnode{n3}{}}
\ncline[offset=.1]{->}{n0}{n1}
\ncline[offset=.1]{->}{n1}{n2}
\ncline[offset=.1]{->}{n2}{n3}
\psrelpoint{n0}{t1}{20}{-4}
\psrelpoint{t1}{t2}{0}{-7}
\rput[l](\x{t1},\y{t1}){Live access paths at entry of block 1:
 $\{ x,\, x\myarrow r,\, x\myarrow r\myarrow r \}$}
\rput[l](\x{t2}, \y{t2}){Corresponding access graph: $G_x^2$}
\psrelpoint{t2}{n0}{50}{0}
\rput(\x{n0},\y{n0}){\rnode{n0}{}}
\psrelpoint{n0}{x0}{7}{0}
\psrelpoint{x0}{n1}{10}{0}
\psrelpoint{n1}{n2}{10}{0}
\rput(\x{x0},\y{x0}){\rnode{x0}{\pscirclebox[framesep=.9]{$x$}}}
\rput(\x{n1},\y{n1}){\rnode{n1}{\pscirclebox[framesep=.2]{$r_1$}}}
\rput(\x{n2},\y{n2}){\rnode{n2}{\pscirclebox[framesep=.2]{$r_2$}}}
\ncline[doubleline=true]{->}{n0}{x0}
\ncline{->}{x0}{n1}
\ncline{->}{n1}{n2}
\rput[l](20,26){\rule{\columnwidth}{.2mm}}
\psrelpoint{base}{n0}{30}{42}
\rput(\x{n0},\y{n0}){\rnode{n0}{}}
\psrelpoint{n0}{n1}{0}{-7}
\rput(\x{n1},\y{n1}){\rnode{n1}{1 \psframebox{$x = x.r$} \white 1}}
\psrelpoint{n1}{n2}{0}{-7}
\rput(\x{n2},\y{n2}){\rnode{n00}{}}
\ncline[offset=.1]{->}{n0}{n1}
\ncline[offset=.1]{->}{n1}{n00}
\ncloop[angleA=270,angleB=90,loopsize=-8,arm=3,offset=3,
  linearc=.1]{->}{n1}{n1}
\psrelpoint{n0}{t1}{20}{-4}
\psrelpoint{t1}{t2}{0}{-7}
\rput[l](\x{t1},\y{t1}){Live access paths at entry of block 1: 
$\{ x,\, x\myarrow r,\, x\myarrow r\myarrow r,\,x\myarrow r\myarrow
  r\myarrow r,\,\ldots \}$}
\rput[l](\x{t2},\y{t2}){Corresponding access graph: $G_x^1$}
\psrelpoint{t2}{n0}{50}{0}
\rput(\x{n0},\y{n0}){\rnode{n0}{}}
\psrelpoint{n0}{x0}{7}{0}
\psrelpoint{x0}{n1}{10}{0}
\rput(\x{x0},\y{x0}){\rnode{x0}{\pscirclebox[framesep=0.9]{$x$}}}
\rput(\x{n1},\y{n1}){\rnode{n1}{\pscirclebox[framesep=0.2]{$r_1$}}}
\ncline[doubleline=true]{->}{n0}{x0}
\ncline{->}{x0}{n1}
\nccurve[nodesepA=-.2mm,nodesepB=-.3mm,angleA=320,angleB=40,
  ncurv=3]{->}{n1}{n1}
\end{pspicture}
\caption{Approximations in Access Graphs}
\label{fig:access.graphs.first}
\rule{\textwidth}{.2mm}
\end{figure}

\subsubsection{Summarization}
\label{sec:summarisation}

Recall that a link is live at a program point~$p$ if it is used along
some control flow path from $p$ to \exitnode.  Since different access
paths may be live along different control flow paths and there may be
infinitely many control flow paths in the case of a loop following
$p$, there may be infinitely many access paths which are live at $p$.
Hence, the lengths of access paths will be unbounded. In such a case
summarization is required.

Summarization is achieved by merging appropriate nodes in access
graphs, retaining all in and out edges of merged nodes.  We explain
merging with the help of
Figure~\ref{fig:access.graphs.first}:
\begin{itemize} 
\item Node $n_1$ in access graph $G_x^1$ indicates references of $n$
  at {\em different execution instances of the same\/} program point.
  Every time this program point is visited during analysis, the same
  state is reached in that the pattern of references after $n_1$ is
  repeated.  Thus all occurrences of $n_1$ are merged into a single
  state.  This creates a cycle which captures the repeating pattern of
  references.
  
\item In $G_x^2$, nodes $n_1$ and $n_2$ indicate referencing $n$ at
  {\em different\/} program points.  Since the references made after
  these program points may be different, $n_1$ and $n_2$ are not
  merged.
\end{itemize}

Summarization captures the pattern of heap traversal in the most
straightforward way.  Traversing a path in the heap requires the
presence of reference assignments \mbox{$\alpha_x = \alpha_y$} such
that $\rho_x$ is a proper prefix of 
$\rho_y$. Assignments in 
Figure~\ref{fig:access.graphs.first} are examples of such
assignments. The structure of the flow of control between such
assignments in a program determines the pattern of heap traversal.
Summarization captures this pattern without the need of control flow
analysis and the resulting structure is reflected in the access graphs
as can be seen in Figure~\ref{fig:access.graphs.first}.  More examples
of the resemblance of program structure and access graph structure can
be seen in the access graphs in Figure~\ref{fig:liveness.info.1}.

\subsubsection{Operations on Access Graphs}
\label{sec:access.graph.operations}

Section~\ref{sec:liveness.specs}  defined liveness  by  applying certain
operations on  access paths.   In this subsection  we define  the corresponding
operations on  access graphs.  Unless  specified otherwise,  the binary
operations  are applied only  to access  graphs having  same root  variable. The
auxiliary operations and associated notations are:

\renewcommand{\graphA}[1]{\mbox{{\sf\em\small G\/}$({#1})$}}
\renewcommand{\graphO}[1]{\mbox{{\sf\em\small\magenta GOnly\/}$({#1})$}}

\begin{itemize} 
\item \RootVar($\rho$)  denotes the root  variable of access path  $\rho$, while
  \RootVar($G$) denotes the root variable of access graph $G$.
\item \FieldName($n$) for a node $n$ denotes the field name component
  of the label of $n$.
\item  \graphA{\rho} constructs access  graphs corresponding  to $\rho$.
  It uses the  current basic block number and the  field names to create
  appropriate  labels for  nodes.  The  instance number  depends  on the
  number    of     occurrences    of    a    field     name    in    the
  block.  \graphA{\rho\myarrow  *} creates  an  access  graph with  root
  variable $x$ and the summary node $n_*$ with an edge from $x$ to $n_*$
  and a self loop over $n_*$.
\item   \lNode{G}  returns   the   last   node  of   a   {\em  linear   graph\/}
$G$ constructed from a given $\rho$.
\item $\clean(G)$ deletes the nodes which are not reachable from the
  entry node. 

\newcommand{\cn}[2]{\mbox{{\sf\em ACN\/}$({#1},{#2})$}}
\item  \cNodes{G}{G'}{S}  computes  the   set  of  nodes  of  $G$  which
  correspond to  the nodes of $G'$  specified  in  the set $S$.
  To compute  \cNodes{G}{G'}{S}, we define \cn{G}{G'}, the  set of pairs
  of {\em all corresponding nodes}.  Let \mbox{$G \equiv \langle n_0, N,
  E\rangle$} and \mbox{$G' \equiv  \langle n'_0, N', E'\rangle$}. A node
  $n$ in $G$ corresponds to a node $n'$ in $G'$ if there there exists an
  access path $\rho$ which is represented by a path from $n_0$ to $n$ in
  $G$ and a path from $n'_0$ to $n'$ in $G'$.

  Formally, \cn{G}{G'} is the least solution of the following  equation:
  \begin{eqnarray*}
     \cn{G}{G'} &=&
       \left\{\begin{array}{llr@{}}
       \emptyset & & \hskip -1cm\RootVar(G) \not= \RootVar(G') \\
       \{\langle n_0, n'_0 \rangle\}
       \cup \{\langle n_j, n'_j \rangle \mid 
       \FieldName(n_j) = \FieldName(n'_j),
       & & \mbox{otherwise}\\
		          {\hskip 3cm} n_i \rightarrow n_j \in E, 
		          n'_i \rightarrow n'_j \in E', \\
			  {\hskip 3cm} \langle n_i, n'_i \rangle\ \in \cn{G}{G'} \}
       \end{array}\right.  \\
       \cNodes{G}{G'}{S} &=& \{ n \mid \langle n, n' \rangle \in \cn{G}{G'}, \,
                          n' \in S \}
  \end{eqnarray*}
\end{itemize}
Note that $\FieldName(n_j) = \FieldName(n'_j)$ would hold even when $n_j$ or $n'_j$ is
the summary node $n_*$.

Let \mbox{$G \equiv \langle n_0, N, E\rangle$} and \mbox{$G' \equiv
\langle n_0, N', E'\rangle$} be access graphs (having the same
entry node). $G$ and $G'$ are equal if $N=N'$ and $E=E'$.

\begin{figure}[t]
{\includegraphics{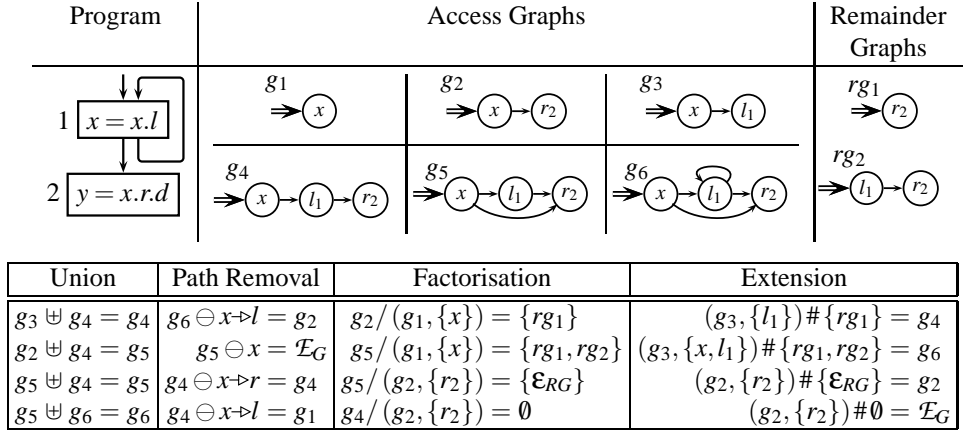}}
\caption{Examples of operations on access graphs.} 
\rule{\textwidth}{.2mm}
\label{fig:exmp.ops.ag}
\end{figure}

 The main operations of interest are defined below and
are illustrated in Figure~\ref{fig:exmp.ops.ag}.

\begin{enumerate}
\item
{\em Union} ($\cupG$).  $G \cupG G'$ combines access graphs $G$ and
$G'$ such that any access path contained in $G$ or $G'$ is contained
in the resulting graph.
\begin{eqnarray*}
  G \cupG G' &=& \left\langle n_0, N \cup N',
   E \cup E'
  \right\rangle
\end{eqnarray*}
The operation $N\cup N'$ treats the nodes with the same label as identical.
Because of associativity, $\cupG$ can be generalized to arbitrary
number of arguments in an obvious manner.
\item

{\em Path Removal} ($\minus$). The operation \mbox{$G\minus\rho$}
removes those access paths in $G$ which have $\rho$ as a prefix.
\begin{eqnarray*}
  G \minus \rho & = & \left\{\begin{array}{ll}
  G & \rho = \Empty \mbox{ or } \RootVar(\rho) \neq \RootVar(G) \\
  \Empty\!_G & \rho \mbox{ is a simple access path} \\
  \clean(\langle n_0, N, E - E_{del} \rangle) \rule{.3cm}{0cm} &
  otherwise
  \end{array} \right.
\end{eqnarray*}
where
\[\begin{array}{rcl}
  E_{del} & = & \{ n_i \rightarrow n_j \mid
                     n_i \rightarrow n_j \in E,
		     n_i \in \cNodes{G}{G^B}{\{\lNode{G^B}\}}, \\
            &   & 
{\white \{ n_i \mathop{\rightarrow}^f n_j \mid}
\FieldName(n_j) = \Front(\rho), 
		     G^B = \graphA{\Base(\rho)}, \\
            &   & 
{\white \{ n_i \mathop{\rightarrow}^f n_j \mid}
		     \uniquepath( G,n_i) \}
  \end{array}\]

\uniquepath($G$, $n$) returns  true if in $G$, all  paths from the entry
node to node $n$ represent the  same access path. Note that path removal
is conservative  in that some paths  having $\rho$ as prefix  may not be
removed. Since  an access graph edge  may be contained in  more than one
access paths,  we have  to ensure  that access paths  which do  not have
$\rho$ as prefix are not erroneously deleted.

\item {\em Factorization} (/).  Recall that the {\em Transfer\/} term in
Definition  \ref{def:explicit.liveness} requires extracting  suffixes of
access  paths  and  attaching  them  to some  other  access  paths.  The
corresponding   operations  on   access  graphs   are   performed  using
factorization and extension.
Given a node \mbox{$m \in (N  - \{n_0\})$} of an access graph $G$, the
{\em Remainder Graph\/} of $G$ at $m$ is the subgraph of $G$ rooted at
$m$ and is denoted by $\subG{G}{m}$. If $m$ does not have any outgoing
edges,  then the  result is  the  empty remainder  graph $\EFG$. 
Let  $M$ be a subset
of the  nodes of $G'$  and $M'$ be  the set of corresponding  nodes in
$G$. Then,  $G/(G',M)$ computes the  set of remainder  graphs of
the successors of nodes in $M'$.
\begin{eqnarray}
  G/(G',M) &=& \{\subG{G}{n_j} \mid n_i \rightarrow n_j \in E, n_i \in
  \cNodes{G}{G'}{M}\}
  \label{eq:factorization}
\end{eqnarray}

A remainder  graph is similar  to an  access graph  except that  (a) its
entry node does not correspond to  a root variable but to a field name
and (b) the  entry node can have incoming edges.  

\item{\em  Extension}. Extending an empty access graph $\Empty\!_G$
results in the empty access graph $\Empty\!_G$. For non-empty graphs,
this operation is defined as follows.
  \begin{enumerate}
  \item{\em Extension with a remainder graph}
    ($\cdot$). 
Let $M$ be a subset of the nodes of $G$ and
    \mbox{$R \equiv \langle n',\, N^{R}, \,E^{R} \rangle$} be a remainder graph.
Then,    \appendG{(G,M)}{R} appends the suffixes in $R$ to the access paths ending
    on nodes in $M$.
    \begin{eqnarray}
      \appendG{(G,M)}{\EFG} &=& G \nonumber \\
      \appendG{(G,M)}{R} &=&
      \left\langle n_0, N \cup N^{R}, 
      E \cup E^{R} \cup 
    \{n_i \rightarrow n' \mid n_i \in M\}
      \right\rangle
  \label{eq:extension.1}
    \end{eqnarray}

  \item {\em Extension with a set of remainder graphs}
    ($\#$). Let $S$ be a set of remainder graphs. Then, \extend{G}{S} extends access 
    graph $G$ with every
    remainder graph in $S$.
    \begin{eqnarray}
      \extend{(G,M)}{\emptyset} &=& \Empty\!_G \nonumber \\
      \extend{(G,M)}{S} &=&
      \displaystyle\mathop{\bigcupG}_{R \in S}\;
      \appendG{(G,M)}{R}
  \label{eq:extension.2}
    \end{eqnarray}
  \end{enumerate}
\end{enumerate}

\subsubsection{Safety of Access Graph Operations}

\begin{figure}[t]
\[
\renewcommand{\arraystretch}{1.2}
\begin{array}{|l|l|l|}
\hline 
\mbox{Operation} &
\mbox{Access Graphs} &
\mbox{Access Paths} 
\\ \hline
\hline 
\mbox{Union} & G_3 = G_1 \cupG\> G_2 
& \AP{G_3,M_3} \supseteq \AP{G_1,M_1} \cup \; \AP{G_2,M_2} 
\\ \hline
\mbox{Path Removal} & G_2 = G_1 \minus\; \rho
& \AP{G_2,M_2} \supseteq 
\AP{G_1,M_1} - \; \{\rho\myarrow\sigma \mid \rho\myarrow\sigma \in \AP{G_1,M_1}\}
\\ \hline
\mbox{Factorization} & S = G_1/(G_2,M)	
& \AP{S,M_s} = \{ \sigma \mid \rho'\myarrow\sigma \in \AP{G_1,M_1},
                \rho' \in \AP{G_2,M} \} 
\\ \hline
\mbox{Extension} & G_2 = \extend{(G_1,M)}{S} &
\AP{G_2,M_2} \supseteq \AP{G_1,M_1} \cup \{ \rho\myarrow\sigma \mid  \rho \in \AP{G_1,M},
                      \sigma \in \AP{S,M_s} \}
\\ \hline
\end{array}
\]
\caption{Safety of Access Graph Operations. $\AP{G,M}$ is the set of
  paths in graph $G$ terminating on nodes in $M$.  
  For graph $G_i$, $M_i$ is the set of all nodes in $G_i$. $S$ is the
  set of remainder graphs and \AP{S,M_s} is the
  set of all paths in all remainder graphs in $S$.}
\label{fig:G.ops.properties}
\rule{\textwidth}{.2mm}
\end{figure}

Since access graphs are not exact representations of sets of access
paths, the safety of approximations needs to be defined
explicitly.  The constraints defined in
Figure~\ref{fig:G.ops.properties} 
capture safety in the context of liveness in the following sense: Every access
path which can possibly be live should be retained by each operation.
Since the complement of liveness is used for nullification, this ensures
that no live access path is considered for nullification.
These properties have been
proved~\cite{hra.AG.Safety} using the PVS 
theorem prover\footnote{Available from \url{http://pvs.csl.sri.com}.}.

\subsection{Data Flow Analysis for Discovering Explicit Liveness}

\label{sec:live-analysis}
For a given root variable \mbox{$v$}, $\Lin{v}(i)$ and $\Lout{v}(i)$
denote the access graphs representing explicitly live access paths at
the entry and exit of basic block $i$.
We use $\Empty\!_G$ as the initial value for $\Lin{v}(i)/\Lout{v}(i)$.

\begin{eqnarray}
\Lin{v}(i) & = & 
\left( \Lout{v}(i) \minus \Lkill{v}(i)\right) \cupG \Lgen{v}(i)
	\label{eq:Elin}
\\
\Lout{v}(i) & = & \left \{ \begin{array}{l@{\ \ \ }l}
		\graphA{v\myarrow *} & i = \exitnode,\; v \in \Global \\ 
		\Empty\!_G & i = \exitnode, \; v \not\in \Global \\
		\displaystyle\mathop{\bigcupG}_{s \in succ(i)}
		\; \Lin{v}(s) & \mbox{otherwise}
		\end{array}\right. 
	\label{eq:Elout}	
\end{eqnarray}
where
\begin{equation*}
\Lgen{v}(i) = \LD{v}(i) \cupG \LT{v}(i)
\end{equation*}

We define $\Lkill{v}(i)$, $\LD{v}(i)$, and $\LT{v}(i)$ depending upon
the statement.

\begin{enumerate}
\item {\em Assignment statement\/} \mbox{$\alpha_x = \alpha_y$}. Apart
from defining the desired terms for $x$ and $y$, we also need to
define them for any other variable $z$. In the following equations,
$G_x$ and $G_y$ denote $\graphA{\rho_x}$ and
$\graphA{\rho_y}$ respectively, whereas $M_x$ and $M_y$ denote
$\lNode{\graphA{\rho_x}}$ and  $\lNode{\graphA{\rho_y}}$ respectively.
\begin{eqnarray}
\LD{x}(i) & \!\!=\!\!& \graphA{\Base(\rho_x)} \nonumber \\
\LD{y}(i) & \!\!=\!\!& \left\{\!\! \renewcommand{\arraystretch}{1.1}
		\begin{array}{l@{\ \ \ }l}
		 \Empty\!_G &  \alpha_y {\mbox{ is {\em New\/} \ldots\ or \NULL}} \\
		\graphA{\Base(\rho_y)}  & \mbox{otherwise}
		\end{array}
		\right.
		\nonumber   \\
\LD{z}(i) & \!\!=\!\!& \Empty\!_G, \mbox{for any variable } z \mbox{ other than } x \mbox{ and }
	y \nonumber  \\
\LT{y}(i) & \!\!=\!\! & \left\{\!\! \renewcommand{\arraystretch}{1.1}
		\begin{array}{l@{\ }l}
		 \Empty\!_G &  \alpha_y \mbox{ is {\em New\/} or \NULL} \\
		\extend{(G_y, M_y)}{} & \mbox{otherwise} \\
		\;\;\;\;\;\;\;\;
		{(\Lout{x}(i)/( G_x, M_x))} 
		\end{array}
		\right.
\label{eq:xfer.y.asgn}\\
\LT{z}(i) & \!\!=\!\! & \Empty\!_G, \mbox{ for any variable } z 
	\mbox{ other than } y \nonumber  \\
\Lkill{x}(i) & \!\!=\!\! & \rho_x \nonumber \\
\Lkill{z}(i) & \!\!=\!\! &\Empty, \mbox{ for any variable } z 
	\mbox{ other than } x \nonumber   
\end{eqnarray}

As stated earlier, the path removal operation deletes an edge only if it
is  contained in  a unique  path. Thus  fewer paths  may be  killed than
desired. This  is a safe approximation.  Another  approximation which is
also  safe is  that only  the  paths rooted  at $x$  are killed.   Since
assignment   to    $\alpha_x$   changes   the    link   represented   by
$\Front(\rho_x)$, for precision, any path which is guaranteed to contain
the link  represented by $\Front(\rho_x)$  should also be  killed.  Such
paths  can be  discovered through  must-alias analysis  which we  do not
perform.

\begin{figure}[t]
\hfill{\includegraphics{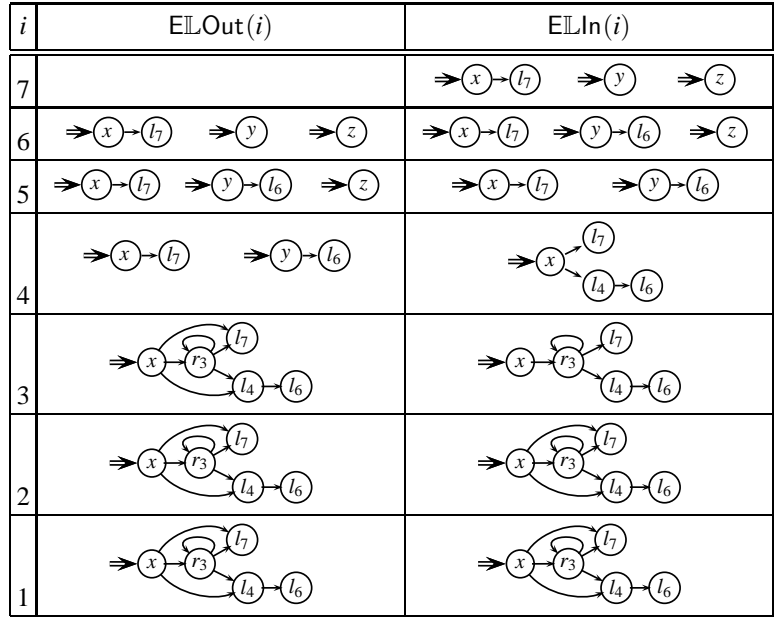}}\hfill\mbox{}
\caption{Explicit liveness for the program in Figure
  \ref{fig:memory.graph} under 
the assumption that all variables are local variables.}
\label{fig:liveness.info.1}
\rule{\textwidth}{.2mm}
\end{figure}

\item  {\em  Function  call\/}  \mbox{$\alpha_x =  f(\alpha_y)$}.   We
  conservatively assume that a function  call may make any access path
  rooted at $y$ or any global reference variable live. 
  Thus this version of our analysis is context insensitive. 
\begin{eqnarray*}
\LD{x}(i) & \!\!=\!\!& \graphA{\Base(\rho_x)}  \\
\LD{y}(i) & \!\!=\!\!&  \graphA{\rho_y\myarrow *} \\
\LD{z}(i) & \!\!=\!\!& \left\{\!\! \renewcommand{\arraystretch}{1.1}
		\begin{array}{l@{\ }l}
	\graphA{z\myarrow *} & \mbox{if $z$ is a global variable}   \\
	\Empty\!_G & \mbox{otherwise}
		\end{array}
		\right.
		\\
\LT{z}(i) & \!\!=\!\! & \Empty\!_G, \mbox{ for all variables } z  \\
\Lkill{x}(i) & \!\!=\!\! & \rho_x \nonumber \\
\Lkill{z}(i) & \!\!=\!\! &\Empty, \mbox{ for any variable } z 
	\mbox{ other than } x \nonumber   
\end{eqnarray*}
\item {\em Return Statement} $\return \; \alpha_x$.
\begin{eqnarray*}
\LD{x}(i) & \!\!=\!\!&  \graphA{\rho_x\myarrow *} \\
\LD{z}(i) & \!\!=\!\!& \left\{\!\! \renewcommand{\arraystretch}{1.1}
		\begin{array}{l@{\ }l}
	\graphA{z\myarrow *} & \mbox{if $z$ is a global variable }   \\
	\Empty\!_G & \mbox{otherwise}
		\end{array}
		\right.
		\\
\LT{z}(i) & \!\!=\!\! & \Empty\!_G, \mbox{ for any variable } z  \\
\Lkill{z}(i) & \!\!=\!\! & \Empty, \mbox{ for any variable } z  
\end{eqnarray*}

\item {\em Use Statements}
\begin{eqnarray*}
\LD{x}(i) & = & \bigcupG\; \graphA{\rho_x} \mbox{ \ for every } 
		\alpha_x.d \mbox{ used in } i \label{eq:direct.use.2} \\
\LD{z}(i) & = & \Empty\!_G \mbox{ \ for any variable $z$ other than $x$ and} y
		\label{eq:direct.use.z} \\
\LT{z}(i) & = & \Empty\!_G, \mbox{ for every variable } z  \label{eq:xfer.z.use} \\
\Lkill{z}(i) & = & \Empty, \mbox{ for every variable } z  \label{eq:kill.z.use} 
\end{eqnarray*}
\end{enumerate}


\begin{example}
\label{exmp:liveness.info.1}
Figure~\ref{fig:liveness.info.1}  lists explicit  liveness information
at different  points of  the program in  Figure \ref{fig:memory.graph}
under the  assumption that all  variables are local  variables.
%
\mybox
\end{example}

Observe that computing liveness using equations~(\ref{eq:Elin}) and (\ref{eq:Elout}) 
results in an MFP (Maximum Fixed Point) solution of data flow analysis whereas
definition~(\ref{def:explicit.liveness}) specifies an MoP (Meet over Paths) solution
of data flow analysis. 
Since the flow functions are non-distributive (see appendix~\ref{sec:non-distributivity}),
the two solutions may be different. 

\section{Other Analyses for Inserting \NULL\ Assignments}
\label{sec:other.analyses}

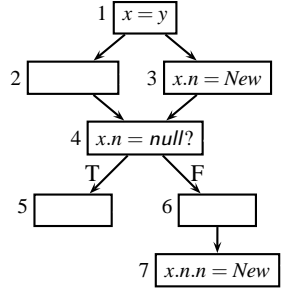
\begin{figure}
\begin{center}
{\psset{unit=.8mm}
\begin{pspicture}(-3,4)(45,50)
\psrelpoint{origin}{n1}{20}{45}
\rput(\x{n1},\y{n1}){\rnode{n1}{1 \psframebox{$x = y$}\white 1\ }}
\psrelpoint{n1}{n2}{-12}{-10}
\rput(\x{n2},\y{n2}){\rnode{n2}{2 \psframebox{\white$x = \mbox{\em New}$}\white 2\ }}
\psrelpoint{n1}{n3}{12}{-10}
\rput(\x{n3},\y{n3}){\rnode{n3}{3 \psframebox{$x.n = \mbox{\em New}$}\white 3\ }}
\psrelpoint{n3}{n4}{-12}{-10}
\rput(\x{n4},\y{n4}){\rnode{n4}{4 \psframebox{$x.n = \NULL?$}\white 4\ }}
\psrelpoint{n4}{n5}{-12}{-12}
\rput(\x{n5},\y{n5}){\rnode{n5}{5 \psframebox{\white$y = x.r$}\white 5\ }}
\psrelpoint{n4}{n6}{12}{-12}
\rput(\x{n6},\y{n6}){\rnode{n6}{6 \white\psframebox{$y = x.r$}\white 6\ }}
\psrelpoint{n6}{n7}{0}{-10}
\rput(\x{n7},\y{n7}){\rnode{n7}{7 \psframebox{$x.n.n = \mbox{\em New}$}\white 7\ }}
\ncline{->}{n1}{n2}
\ncline{->}{n1}{n3}
\ncline{->}{n2}{n4}
\ncline{->}{n3}{n4}
\ncline{->}{n4}{n5}
\bput[0pt](.7){\small T}
\ncline{->}{n4}{n6}
\aput[0pt](.7){\small F}
\ncline{->}{n6}{n7}
\end{pspicture}
}
\end{center}
\caption{Explicit liveness information is not sufficient for nullification.}

\label{fig:pgm.nullasgn}
\rule{\textwidth}{.2mm}
\end{figure}

Explicit liveness alone is not  enough to decide whether an assignment
\mbox{$\alpha_x = \NULL$}  can be safely inserted at  $p$.  We have to
additionally ensure that:

\begin{itemize}
\item $\Front(\rho_x)$ is not live through an alias created before the
program  point $p$.  The extensions  required to  find all  live access
paths, including those created due  to aliases, is discussed in section
\ref{sec:aliasing.def}.
\item Dereferencing links  during the
execution of the inserted statement \mbox{$\alpha_x = \NULL$} does not
cause an exception.  This is done through {\em  availability} and {\em
anticipability}    analysis   and    is    described   in    section
\ref{sec:availability.def}.
\end{itemize}

Both these requirements are illustrated through the example shown
below:

\begin{example}
\label{exmp.illegal.dereference}
In   Figure~\ref{fig:pgm.nullasgn},  access  path \mbox{$y\myarrow n$}  
is not explicitly live in block 6. However,  \mbox{$\Front(y\myarrow n)$} and 
\mbox{$\Front(x\myarrow n)$} represent the same link due to the 
assignment \mbox{$x=y$}. Thus \mbox{$y\myarrow n$} is implicitly live
and setting it to \NULL\ in block 6 will raise an exception in block 7. 
Also, \mbox{$x\myarrow n\myarrow n$}  
is not live  in block 2.  However, it cannot be  set to
\NULL\ since  the object pointed  to by \mbox{$x\myarrow n$}  does not
exist  in  memory when  the  execution  reaches  block 2.   Therefore,
insertion of \mbox{$x.n.n = \NULL$} in block 2 will raise an exception
at run-time.  \mybox
\end{example}

\subsection{Computing Live Access Paths}
\label{sec:aliasing.def}

Recall that an access path is  live if it is either explicitly live or
  shares its \Front\ with some  explicitly live path.  The property of
  sharing is  captured by {\em  aliasing}.  Two access  paths $\rho_x$
  and  $\rho_y$  are  {\em   aliased\/}  at  a  program  point~$p$  if
  $\Target(\rho_x)$ is  same as  $\Target(\rho_y)$ at $p$  during some
  execution of  the program.  They  are {\em link-aliased\/}  if their
  frontiers represent the same  link; they are {\em node-aliased\/} if
  they are aliased but their frontiers do not represent the same link.
  Link-aliases   can   be   derived   from  node-aliases   (or   other
  link-aliases)  by adding  the  same field  names  to aliased  access
  paths.

Alias  information  is {\em  flow-sensitive\/}  if  the  aliases at  a
program  point  depend on  the  statements  along  control flow  paths
reaching  the point.  Otherwise  it is  flow insensitive.   Among flow
sensitive aliases, two access paths are {\em must-aliased\/} at $p$ if
they are aliased along every  control flow path reaching $p$; they are
{\em may-aliased\/} if  they are aliased along some  control flow path
reaching  $p$.   As   an  example,  in  Figure~\ref{fig:memory.graph},
\mbox{$x\myarrow\lptr$}   and    \mbox{$y$}   are   must-node-aliases,
\mbox{$x\myarrow\lptr\myarrow\lptr$}  and  \mbox{$y\myarrow\lptr$} are
must-link-aliases, and $w$ and $x$ are node-aliases at line 5.

We compute flow sensitive may-aliases (without kills) using the algorithm described by
\citeN{hind99interprocedural}  and  use  pairs  of  access  graphs  for
compact  representation of  aliases.  Liveness  is computed  through a
backward  propagation much  in the  same manner  as  explicit liveness
except that it is ensured that the live paths at each program point is
closed under may-aliasing. This  requires the following two changes in
the earlier  scheme.
\begin{enumerate}
\item {\em Inclusion of Intermediate Nodes in Access Graphs\/}. Unlike
explicit liveness, live  access paths may not be  prefix closed.  This
is because the frontier of a live access path $\rho_x$ may be accessed
using  some  other  access  path  and  not  through  the  links  which
constitute $\rho_x$. Hence prefixes of $\rho_x$ may not be live. In an
access graph  representing liveness, all paths may  not represent live
links.  We  therefore modify the access  graph so that  such paths are
not described by the access  graph. In order to make this distinction,
we  divide the  nodes  in an  access  graphs in  two categories:  {\em
final\/} and {\em intermediate\/}.  The only access paths described by
the  access  graph  are  those  which end  at  final  nodes.
\footnote{These  two  categories  are  completely  orthogonal  to  the
labeling  criterion of  the nodes.}   
This change  affects  the access
graph operations in the following manner:
      \begin{itemize}
      \item The equality of graphs now must consider equality of the 
            sets of intermediate nodes and the sets of final nodes
            separately.
      \item Graph  constructor \graphA{\rho_x} marks all  nodes in the
            resulting  graph  as  final  implying that  all  non-empty
            prefixes  of  $\rho_x$ are  contained  in  the graph.   We
            define  a new  constructor \newGraphO{\rho_x}  which marks
            only  the  last node  as  final  and  all other  nodes  as
            intermediate implying  that only $\rho_x$  is contained in
            the graph.
      \item Whenever multiple nodes with identical labels are combined, if any 
            instance of
            the node is final then the resulting node is treated as final. This 
	    influences union $(\!\!\cupG\!\!)$ and extension (\extend{}{}).
	\item The set $M$ used in defining factorization and extension 
	(equations~\ref{eq:factorization}, \ref{eq:extension.1}, \ref{eq:extension.2})
	and the safety properties of access graph operations 
	(Figure~\ref{fig:G.ops.properties})
	contain final nodes only.
	\item Extension $\appendG{G}{RG}$ marks all nodes in $G$ as intermediate.
	      If $G$ and $RG$ have a common node then the status of the node is 
	       governed by its status in $RG$.
        \item The $\clean(G)$ operation is modified to delete those intermediate nodes
              which do not have a path leading to a final node.
      \end{itemize}
\item {\em Link Alias Closure\/}. To  discover all  link  aliases of  a  
      live access  we compute link alias closure as defined below.
Given an alias set \Aset, the set of link aliases of an access path 
$\rho_x\myarrow f$ is the least solution of:
\begin{equation*}
\LnA(\rho_x\myarrow f,\Aset)  = \{ \rho_y\myarrow f \mid 
	\langle \rho_x,\rho_y\rangle \in \Aset \mbox{ or }
	\langle \rho_x,\rho_y\rangle \in \LnA(\rho_x,\Aset) 
	\}
\end{equation*}

 Given an alias pair $ \langle g_x,g_y\rangle$
link aliases of $G_x$ rooted at $y$ are included
in the access graph $G_y$ as follows:
\begin{equation}
  \LnG(G_y,G_x, \langle g_x,g_y\rangle)= G_y  \cupG
\extend{(g_y,m_y)}{((G_x/(g_x,m_x)}) \!-\!\EFG)
	\label{eq:link.alias.computation} 
\end{equation}
where $m_y$ and $m_x$ are the singleton sets containing the final nodes of
$g_y$ and $g_x$ respectively.
$\EFG$ has to be removed from set of remainder graphs because we want
to transfer non-empty links only. 
Complete liveness is computed as the least solution of the following
equations 
\begin{eqnarray*}
\CLin{v}(i) & = & \TLin{v}(i) 
	\displaystyle\mathop{\bigcupG}_{\raisebox{.1mm}{$\scriptstyle
		\langle g_v,g_u\rangle$} 
		\in 
		\raisebox{-.2mm}{\NodeAin(i)}}
\LnG\left(\CLin{v}(i),\CLin{u}(i), \langle g_v,g_u\rangle \right)
\\
\CLout{v}(i) & = & \left \{ \begin{array}{l@{\ \ \ }l}
	\graphA{v\myarrow *}
                & i = \exitnode, \; v \in \Global \\
                &\mbox{or }  v \in \param\\
\Empty\!_G \cupG 
	\LnG\left( \CLout{v}(i),
                   \CLout{u}(i), \langle g_v,g_u\rangle 
           \right)
		& i = \exitnode,\; v \not\in \Global,
 \\ 
		& \langle g_v,g_u\rangle \in \NodeAout(i)
\\
		\displaystyle\mathop{\bigcupG}_{s \in succ(i)}
		\; \CLin{v}(s) & \mbox{otherwise}
		\end{array}\right. 
\end{eqnarray*}
where $\TLin{v}(i)$ is same as $\Lin{v}(i)$
except that $\Lout{v}(i)$ is replaced by
$\CLout{v}(i)$ in the main equation (equation \ref{eq:Elin}) and in
the computation of {\em Transfer\/} (equation \ref{eq:xfer.y.asgn}).

\end{enumerate}

\begin{figure}[t]
\newcommand{\pair}[2]{\mbox{$\left\langle \raisebox{-1.3mm}{#1}, 
	\raisebox{-1.3mm}{#2}\right\rangle$}}
\newcommand{\Gx}
{\scalebox{.95}{
{\psset{unit=.8mm}
		\begin{pspicture}(1,.25)(9,5.5)
		\putnode{a}{origin}{0}{3}{}
		\putnode{x}{a}{8}{0}{\pscirclebox{$\,x\,$}}
		\ncline[doubleline=true]{->}{a}{x}
		\end{pspicture}
		}
}
}
\newcommand{\Gxr}
{\scalebox{.95}{
{\psset{unit=.8mm}
		\begin{pspicture}(1,.5)(17.5,5.5)
		\putnode{a}{origin}{0}{3}{}
		\putnode{x}{a}{8}{0}{\pscirclebox[linestyle=dashed, dash=.6 .6]{$x$}}
		\putnode{r}{x}{9}{0}{\pscirclebox[framesep=.8]{$r_3$}}
		\ncline[doubleline=true]{->}{a}{x}
		\ncline{->}{x}{r}
		\end{pspicture}
		}
}
}
\newcommand{\Gxl}
{\scalebox{.95}{
{\psset{unit=.8mm}
		\begin{pspicture}(1,.5)(17.5,5.5)
		\putnode{a}{origin}{0}{3}{}
		\putnode{x}{a}{8}{0}{\pscirclebox[linestyle=dashed, dash=.6 .6]{$x$}}
		\putnode{r}{x}{9}{0}{\pscirclebox[framesep=.7]{$l_4$}}
		\ncline[doubleline=true]{->}{a}{x}
		\ncline{->}{x}{r}
		\end{pspicture}
		}
}
}
\newcommand{\Gw}
{\scalebox{.95}{
{\psset{unit=.8mm}
		\begin{pspicture}(1,.5)(9,5.5)
		\putnode{a}{origin}{0}{3}{}
		\putnode{x}{a}{8}{0}{\pscirclebox{$w$}}
		\ncline[doubleline=true]{->}{a}{x}
		\end{pspicture}
		}
}
}
\newcommand{\Gy}
{\scalebox{.95}{
{\psset{unit=.8mm}
		\begin{pspicture}(1,.5)(9,5.5)
		\putnode{a}{origin}{0}{3}{}
		\putnode{x}{a}{8}{0}{\pscirclebox{$y$}}
		\ncline[doubleline=true]{->}{a}{x}
		\end{pspicture}
		}
}
}
\newcommand{\Gyl}
{\scalebox{.95}{
{\psset{unit=.8mm}
		\begin{pspicture}(1,.5)(17.5,5.5)
		\putnode{a}{origin}{0}{3}{}
		\putnode{x}{a}{8}{0}{\pscirclebox[linestyle=dashed, dash=.6 .6]{$y$}}
		\putnode{r}{x}{9}{0}{\pscirclebox[framesep=.7]{$l_6$}}
		\ncline[doubleline=true]{->}{a}{x}
		\ncline{->}{x}{r}
		\end{pspicture}
		}
}
}

\[
\begin{array}{|c|l|l|} \hline
i & \multicolumn{1}{|c|}{\NodeAin(i)}  & \multicolumn{1}{|c|}{\NodeAout(i)} \\ \hline
\hline
\rule[-.75em]{0em}{2em}%
1 & \multicolumn{1}{|c|}{\emptyset}          & \pair{\Gx}{\Gw} \\\hline
\rule[-.75em]{0em}{2em}%
2 & \pair{\Gx}{\Gw}, \pair{\Gx}{\Gxr}
  & \pair{\Gx}{\Gw}, \pair{\Gx}{\Gxr} \\\hline
\rule[-.75em]{0em}{2em}%
3 & \pair{\Gx}{\Gw}, \pair{\Gx}{\Gxr}
  & \pair{\Gx}{\Gw}, \pair{\Gx}{\Gxr} \\\hline
\rule[-.75em]{0em}{2em}%
4 & \pair{\Gx}{\Gw}, \pair{\Gx}{\Gxr}
  & \pair{\Gx}{\Gw}, \pair{\Gx}{\Gxr}, \\
\rule[-.75em]{0em}{2em}%
 & &  \pair{\Gy}{\Gxl} \\\hline
\rule[-.75em]{0em}{2em}%
5 & \pair{\Gx}{\Gw}, \pair{\Gx}{\Gxr}, 
  & \pair{\Gx}{\Gw}, \pair{\Gx}{\Gxr}\\
\rule[-.75em]{0em}{2em}%
 & \; \pair{\Gy}{\Gxl}
  &  \;\pair{\Gy}{\Gxl} \\\hline
\rule[-.75em]{0em}{2em}%
6 & \pair{\Gx}{\Gw}, \pair{\Gx}{\Gxr},
  & \pair{\Gx}{\Gw}, \pair{\Gx}{\Gxr},\\
\rule[-.75em]{0em}{2em}%
 &  \;\pair{\Gy}{\Gxl}
  &  \;\pair{\Gy}{\Gxl}, \pair{\Gy}{\Gyl} \\\hline
\rule[-.75em]{0em}{2em}%
7 & \pair{\Gx}{\Gw}, \pair{\Gx}{\Gxr},
  & \pair{\Gx}{\Gw}, \pair{\Gx}{\Gxr},\\
\rule[-.75em]{0em}{2em}%
 &  \;\pair{\Gy}{\Gxl}, \pair{\Gy}{\Gyl}
  &  \;\pair{\Gy}{\Gxl}, \pair{\Gy}{\Gyl} \\\hline
\end{array}
\]

\caption{Alias pairs for the running example from 
	Figure\protect~(\ref{fig:memory.graph}). }
\end{figure}

\begin{figure}[t]
\begin{center}
\includegraphics{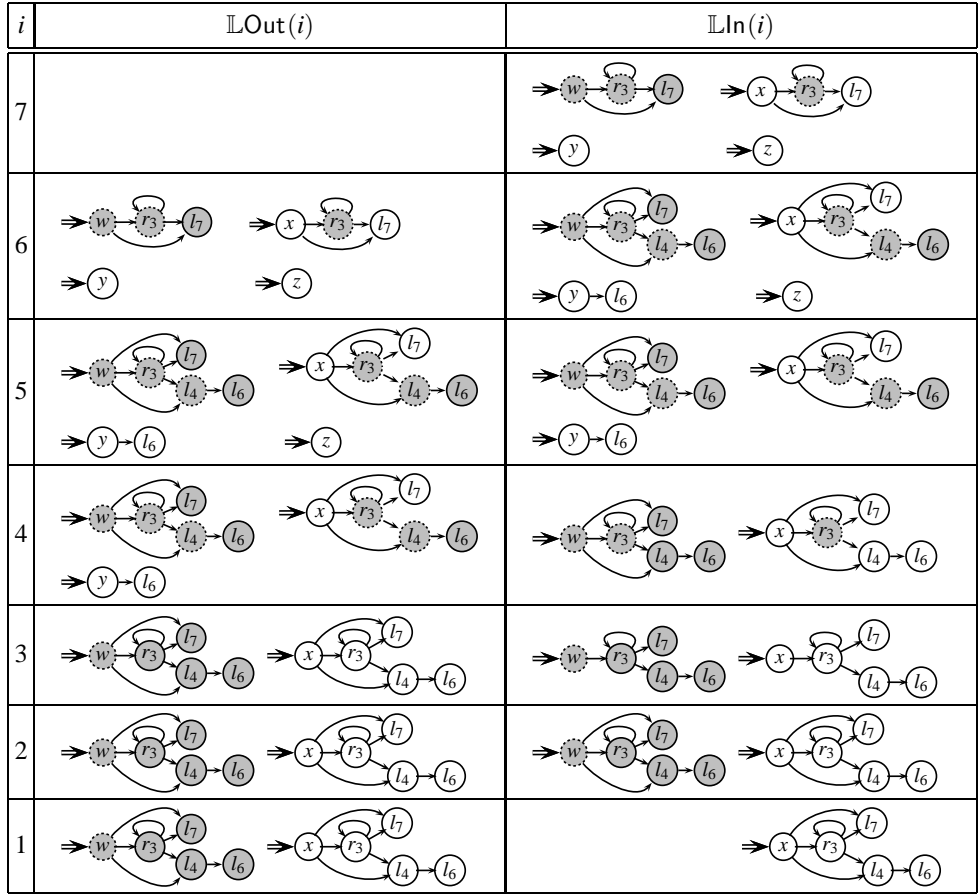}
\end{center}

\caption{Liveness access graphs including implicit liveness
  information for the program in Figure~\ref{fig:memory.graph}.
Gray nodes are nodes included by link-alias computation. Intermediate nodes
are shown with dotted lines.
}
\label{fig:link-alias.info}
\rule{\textwidth}{.2mm}

\end{figure}

\begin{example}
\label{exmp.alias.info}
Figure~\ref{fig:pgm.nullasgn} shows the may-alias information for our
running example from Figure\protect~\ref{fig:memory.graph}. Observe that
the access graphs used for storing alias information have only the last node
as final and all other nodes as intermediate. 
Figure~\ref{fig:link-alias.info} shows the liveness access graphs augmented
with implicit liveness.  \mybox
\end{example}

Observe that in the presence of cyclic data structures, we will get alias pairs of
the form \mbox{$\langle \rho,\rho\myarrow\sigma\rangle$}. If a link in the cycle is 
live then the link alias closure will ensure that all possible links are marked live by creating
cycles in the access graphs. This may cause approximation but would be safe.

\subsection{Availability and Anticipability of Access Paths}
\label{sec:availability.def}

Example~\ref{exmp.illegal.dereference} shows  that safety of inserting
an  assignment  \mbox{$\alpha_x  =  \NULL$}  at  a  program  point~$p$
requires  that   whenever  control   reaches  $p$,  every   prefix  of
$\Base(\rho_x)$ has a non-\NULL\ l-value.  Such an access path is said
to be {\em  accessible} at $p$.  Our use  of accessibility ensures the
preservation  of  semantics  in   the  following  sense:  Consider  an
execution path  which does not  have a dereferencing exception  in the
unoptimized  program.  Then  the proposed  optimization will  also not
have any dereferencing exception in the same execution path.

\subsubsection{Defining Availability and Anticipability}

We define  an access path $\rho_x$ to  be accessible at $p$  if all of
its prefixes are  {\em available} or {\em anticipable} at $p$:
\begin{itemize}
\item  An access  path  $\rho_x$  is {\em  available\/}  at a  program
  point~$p$, if along every path  reaching $p$, there exists a program
  point~$p'$  such  that $\Front(\rho_x)$  is  either dereferenced  or
  assigned a non-\NULL\ l-value at $p'$ and is not made \NULL\ between
  $p'$ to $p$.
\item  An access  path $\rho_x$  is {\em  anticipable\/} at  a program
  point~$p$, if  along every path starting  from $p$, $\Front(\rho_x)$
  is dereferenced before being assigned.
\end{itemize}
Since both these properties are {\em all paths\/} properties, all
may-link aliases of the left hand side of an assignment need to be killed.
Conversely, these properties can be made more precise by including
must-aliases in the set of anticipable or available paths.

Recall that  comparisons in conditionals consists  of simple variables
only.   The   use   of   these   variables  does   not   involve   any
dereferencing.  Hence a  comparison $x  == y$  does not  contribute to
accessibility of $x$ or $y$. 

\begin{figure}[t]
\begin{center}
\scalebox{1.2}{%
$
\begin{array}{|@{\ }l@{\ }|@{\ }c@{\ }|@{\ }c@{\ }|@{\ }c@{\ }|}
\hline
\mbox{Statement } s & \AVK_s & \AVD_s & \AVT_s(X) \\  \hline\hline
\alpha_x = \alpha_y & 
	\{ \rho_z\myarrow * \mid  \rho_z \in \ \LnA(\rho_x,\NodeAin(s)) \} & 
	\prefix(\Base(\rho_x)) &
	\{ \rho_x\myarrow \sigma \mid \rho_y\myarrow \sigma \in X\} 
	\\
& & \cup \prefix(\Base(\rho_y))  &
 		\\ \hline
 \alpha_x = f(\alpha_y) & 
	\{ \rho_z\myarrow * \mid  \rho_z \in \ \LnA(\rho_x,\NodeAin(s)) \} & 
	\prefix(\Base(\rho_x))
 & \emptyset \\ 
 		 \hline
\alpha_x = \new & 
	\{ \rho_z\myarrow * \mid  \rho_z \in \ \LnA(\rho_x,\NodeAin(s)) \} & 
	\prefix(\rho_x) & \emptyset
 		\\ \hline
\alpha_x = \NULL & 
	\{ \rho_z\myarrow * \mid  \rho_z \in \ \LnA(\rho_x,\NodeAin(s)) \} & 
	\prefix(\Base(\rho_x)) & \emptyset
 		\\ \hline
\use  \; \alpha_y.d  & \emptyset & \prefix(\rho_y) & \emptyset
 		\\ \hline
\return \;  \alpha_y & \emptyset &  \prefix(\Base(\rho_y))
& \emptyset 
 		\\ \hline
\mbox{other}  & \emptyset &  \emptyset & \emptyset 
 		\\ \hline
\end{array}
$}
\end{center}
\caption{Flow functions for availability. $\NodeAin(s)$ denotes the set of
may-aliases at the entry of $s$.}
\label{fig:flow.fun.availability}
\rule{\textwidth}{.2mm}
\end{figure}

\begin{definition}{\rm\bf Availability}.
\label{def:availability}
The set of paths which are available at a program point~$p$,
denoted by $\avail_{p}$, is defined as follows.
  \begin{eqnarray*}
    \avail_p& = & 
    \displaystyle\bigcap_{\onepath \in Paths(p)}(\Pavail^{\onepath}_{p}) 
  \end{eqnarray*}
where, $\onepath \in Paths(p)$ is a control flow path \entrynode\ to $p$ and
$\Pavail^{\onepath}_{p}$ denotes the
availability at $p$ along $\onepath$ and is defined as follows. 
If $p$ is not \entrynode\ of the procedure being analyzed, then let the statement 
which precedes it be denoted by
  $s$ and the program point immediately preceding $s$ be denoted by $p'$. Then,
  \begin{eqnarray*}
    \Pavail^{\onepath}_{p}& = & \left\{\begin{array}{cl}
    \emptyset & p \mbox{ is  \entrynode} \\
    \Savail_{s}(\Pavail^{\onepath}_{p'})  & \mbox{otherwise}
    \end{array}\right. 
  \end{eqnarray*}
where the flow function for $s$ is defined as follows:
  \begin{eqnarray*}
    \Savail_{s}(X) & = & (X - \AVK_s) \; \cup \; \AVD_s \;\cup\;\AVT_s(X) 
  \end{eqnarray*}
$\AVK_s$ denotes the sets of access paths which cease to be available after 
statement $s$, $\AVD_s$ denotes the set of access paths which become available due 
to local effect of $s$ and $\AVT_s(X)$ denotes the
the set of access paths which become available after $s$ due to transfer. 
They are defined in Figure~\ref{fig:flow.fun.availability}. \mybox
\end{definition}

In a similar manner, we define anticipability of access paths.

\begin{figure}[t]
\begin{center}
\scalebox{1.2}{%
$
\begin{array}{|@{\ }l@{\ }|@{\ }c@{\ }|@{\ }c@{\ }|@{\ }c@{\ }|}
\hline
\mbox{Statement } s & \ANTK_s & \ANTD_s & \ANTT_s(X) \\  \hline\hline
\alpha_x = \alpha_y & 
	\{ \rho_z\myarrow * \mid  \rho_z \in \ \LnA(\rho_x,\NodeAout(s)) \} & 
	\prefix(\Base(\rho_x))  &
	\{ \rho_y\myarrow \sigma \mid \rho_x\myarrow \sigma \in X\} 
	\\
& & \cup \prefix(\Base(\rho_y))  &
 		\\ \hline
 \alpha_x = f(\alpha_y) & 
	\{ \rho_z\myarrow * \mid  \rho_z \in \ \LnA(\rho_x,\NodeAout(s)) \} & 
	\prefix(\Base(\rho_x))
 & \emptyset \\ 
& & \cup \prefix(\Base(\rho_y))  &
		\\ 	 \hline
\alpha_x = \new & 
	\{ \rho_z\myarrow * \mid  \rho_z \in \ \LnA(\rho_x,\NodeAout(s)) \} & 
	\prefix(\Base(\rho_x)) & \emptyset
 		\\ \hline
\alpha_x = \NULL & 
	\{ \rho_z\myarrow * \mid  \rho_z \in \ \LnA(\rho_x,\NodeAout(s)) \} & 
	\prefix(\Base(\rho_x)) & \emptyset
 		\\ \hline
\use  \; \alpha_y.d  & \emptyset & \prefix(\rho_y) & \emptyset
 		\\ \hline
\return \;  \alpha_y & \emptyset &  \prefix(\Base(\rho_y))
& \emptyset 
 		\\ \hline
\mbox{other}  & \emptyset &  \emptyset & \emptyset 
 		\\ \hline
\end{array}
$}
\end{center}
\caption{Flow functions for anticipability. $\NodeAout(s)$ denotes the set of
may-aliases at the exit of $s$.}
\label{fig:flow.fun.anticipability}
\rule{\textwidth}{.2mm}
\end{figure}

\begin{definition}{\rm\bf Anticipability}.
\label{def:anticipability}
The set of paths which are anticipable at a program point~$p$,
denoted by $\ant_p$ is defined as follows.
  \begin{eqnarray*}
    \ant_p& = & 
    \displaystyle\bigcap_{\onepath \in Paths(p)}(\Pant^{\onepath}_{p}) 
  \end{eqnarray*}
where, $\onepath \in Paths(p)$ is a control flow path $p$ to \exitnode\ and
$\Pavail^{\onepath}_{p}$ denotes the
anticipability at $p$ along $\onepath$ and is defined as follows. 
If $p$ is \exitnode\ then let the statement which follows it be denoted by
  $s$ and the program point immediately following $s$ be denoted by $p'$. Then,
  \begin{eqnarray*}
    \Pant^{\onepath}_{p}& = & \left\{\begin{array}{cl}
    \emptyset & p \mbox{ is  \exitnode} \\\
    \Sant_{s}(\Pant^{\onepath}_{p'})  & \mbox{otherwise}
    \end{array}\right. 
  \end{eqnarray*}
where the flow function for $s$ is defined as follows:
  \begin{eqnarray*}
    \Sant_{s}(X) & = & (X - \ANTK_s) \; \cup \; \ANTD_s \;\cup\;\ANTT_s(X) 
  \end{eqnarray*}
$\ANTK_s$ denotes the sets of access paths which cease to be anticipable before 
statement $s$, $\ANTD_s$ denotes the set of access paths which become anticipable due 
to local effect of $s$ and $\ANTT_s(X)$ denotes the
the set of access paths which become anticipable before $s$ due to transfer.
They are defined in Figure~\ref{fig:flow.fun.anticipability}. \mybox
\end{definition}

Observe that both $\avail_{p}$ and $\ant_{p}$ are prefix-closed.

\subsubsection{Data Flow Analyses for Availability and Anticipability}

Availability and Anticipability are {\em all (control-flow) paths\/}
properties in that the desired property must hold along every path
reaching/leaving the program point under consideration.  Thus these
analyses identify access paths which are common to all control flow
paths {\em including acyclic control flow paths}.  Since acyclic
control flow paths can generate only acyclic\footnote{In the presence
of cycles in heap, considering only acyclic access paths results in
an approximation which is safe for availability and anticipability.} and hence
finite access paths, anticipability and availability analyses deal
with a finite number of access paths and summarization is not
required.

Thus  there is  no  need to  use  access graphs  for availability  and
anticipability analyses. The data flow analysis can be performed using
a set  of access paths  because the access  paths are bounded  and the
sets would be finite.  Moreover, since the access paths resulting from
anticipability and availability are  prefix-closed, they can be
represented efficiently.

The data flow equations are same as the definitions of these analyses
except that definitions are path-based 
{(i.e. they define MoP solution)}
while the data flow equations are edge-based 
{(i.e. they define MFP solution)}
as is customary in data flow analysis. In other words,
the data flow information is merged at the intermediate 
points and availability and anticipability information is derived
from the corresponding information at the preceding and following program
point respectively. 
{As observed in appendix~\ref{sec:non-distributivity},
the flow functions in availability and anticipability analyses are
non-distributive hence MoP and MFP solutions may be different.
}

For brevity, we omit the data flow equations.
We use the universal set of
access paths as the initial value for all blocks other than
\entrynode\ for availability analysis and \exitnode\ for
anticipability analysis.

\begin{figure}[t]
\hfill{\includegraphics{fig-av-ant.epsi}}\hfill\mbox{}
\caption{Availability and anticipability for the program in Figure
  \ref{fig:memory.graph}.}
\label{fig:av-ant-info}
\rule{\textwidth}{.2mm}
\end{figure}

\begin{example} \label{exmp:av-ant-info}
  Figure~\ref{fig:av-ant-info} gives the availability and
  anticipability information for program in Figure~\ref{fig:memory.graph}.
$\Avin{}(i)$ and $\Avout{}(i)$ denote the set
of available access paths before and after the statement $i$, while
$\Antin{}(i)$ and $\Antout{}(i)$ denote the set of anticipable access
paths before and after the statement $i$. 
%
\mybox
\end{example}

\section{\NULL\ Assignment Insertion} 
\label{sec:nullability}

We now explain how the analyses described in preceding sections can be
used to insert appropriate \NULL\ assignments to nullify dead links.
The inserted assignments should be safe and profitable as defined below.

\begin{definition}{\rm\bf Safety}.
\label{def:safety}
It is safe to insert an assignment \mbox{$\alpha = \NULL$} at a program
point~$p$ if and only if $\rho$ is not live at $p$ and $\Base(\rho)$ can
be dereferenced without raising an exception.
\end{definition}

An access path $\rho$ is {\em nullable} at a program point~$p$ if and
only if it is safe to insert assignment \mbox{$\alpha = \NULL$} at
$p$.

\begin{definition}{\rm\bf Profitability}.  
\label{def:profitability}
It is profitable to insert an assignment \mbox{$\alpha = \NULL$} at a
program point~$p$ if and only if no proper prefix of $\rho$ is nullable
at $p$ and the link corresponding to $\Front(\rho)$ is not made \NULL\
before execution reaches $p$.
\end{definition}

Note that profitability definition is  strict in that every control flow
path  may  nullify  a  particular  link  only  once.   Redundant  \NULL\
assignments on any  path are prohibited.  Since control  flow paths have
common segments,  a \NULL\ assignment may  be partially redundant
in the sense that it may be  redundant along one path but not along some
other  path.  Such  \NULL\ assignments  will be  deemed  unprofitable by
Definition~\ref{def:profitability}.   Our algorithm may  not be  able to
avoid all redundant assignments.

\begin{example}
We  illustrate  some situations  of  safety  and  profitability for  the
program in Figure~\ref{fig:memory.graph}.
\begin{itemize}
\item Access path \mbox{$x\myarrow\lptr\myarrow\lptr$} is not nullable
  at the entry of 6. This is because
  \mbox{$x\myarrow\lptr\myarrow\lptr$} is implicitly live, due to the
  use of \mbox{$y\myarrow\lptr$} in 6. Hence it is not safe to insert
  \mbox{$x.\lptr.\lptr = \NULL$} at the entry of 6.
\item Access path \mbox{$x\myarrow\rptr$} is nullable at the entry of
  4, and continues to be so on the path from the entry of 4 to the
  entry of 7.  The assignment \mbox{$x.\rptr=\NULL$} is profitable
  only at the entry of 4.
%
\mybox
\end{itemize}
\end{example}

\newcommand{\LIVE}[1]{\mbox{{\em Live\/}($#1$)}}
\newcommand{\RCH}[1]{\mbox{{\em Available\/}($#1$) $\cup$ {\em Anticipable\/}($#1$)}}

Section~\ref{sec:null.criteria} describes the criteria for deciding
whether a given path $\rho$ should be considered for a \NULL\
assignment at a program point $p$.  Section~\ref{sec:null.candidates}
describes how we create the set of candidate access paths.
Let    \LIVE{p},    \mbox{\em    Available\/}($p$),   and    \mbox{\em
Anticipable\/}($p$)  denote set of live paths,
set  of available  paths and  set of  anticipable paths  respectively at
program point $p$.\footnote{Because availability and anticipability properties are
  prefix closed, $\Base(\rho)\in\RCH{p}$ guarantees that all proper
  prefixes of $\rho$ are either available or anticipable.}
They refer to
$\CLin{}(i)$, $\Avin{}(i)$, and $\Antin{}(i)$ respectively when $p$ is \In{i}.
When $p$ is \Out{i}, they refer to $\CLout{}(i)$, $\Avout{}(i)$, and 
$\Antout{}(i)$ respectively. 

\subsection{Computing Safety and Profitability}
\label{sec:null.criteria} 

To find out if $\rho$ can be nullified at $p$, we compute two
predicates: \cannullify\ and \nullify. $\cannullify(\rho, p)$ captures
the safety property---it is true if insertion of assignment
\mbox{$\alpha = \NULL$} at program point $p$ is safe.
\begin{eqnarray}
  \cannullify(\rho, p) & = & 
  \rho\not\in\LIVE{p}\; \wedge\; \Base(\rho)\in\RCH{p}
  \label{eq:cannullify}
\end{eqnarray}

$\nullify(\rho, p)$ captures the  profitability property---it is true if
insertion of assignment \mbox{$\alpha =  \NULL$} at program point $p$ is
profitable. To compute  \nullify, we note that it  is most profitable to
set  a link  to  \NULL\ at  the earliest  point  where it  ceases to  be
live.  Therefore,
the  \nullify\  predicate  at a  point  has  to  take into  account  the
possibility of  \NULL\ assignment insertion at previous  point(s). For a
statement  $i$ in  the program,  let $\In{i}$  and $\Out{i}$  denote the
program points immediately before and after $i$. Then,
\begin{eqnarray}
  \nullify(\rho, \Out{i}) &=& \cannullify(\rho, \Out{i})
  \wedge(\bigwedge_{\rho' \in \mbox{\scriptsize\em ProperPrefix}(\rho)}
  \rho' \not\in\LIVE{\Out{i}})
  \nonumber \\
  &&\wedge\; (\neg \cannullify(\rho, \In{i}) \vee
  \neg \Transp(\rho, i))
  \label{eq:nullify:A}\\
  \nullify(\rho, \In{i}) &=& \cannullify(\rho, \In{i})
  \wedge(\bigwedge_{\rho' \in \mbox{\scriptsize\em ProperPrefix}(\rho)}
  \rho' \not\in\LIVE{\In{i}})
  \nonumber \\
  && \wedge\; \rho\neq\Lhs(i) \wedge
  (\neg\!\!\!\! \bigwedge_{j \in pred(i)}\!\!\!\!\!\!
  \cannullify(\rho, \Out{j}))
  \label{eq:nullify:B}
\end{eqnarray}
where, \Transp($\rho$, $i$)  denotes that $\rho$ is transparent  with respect to
statement $i$, i.e.  no prefix of $\rho$ is may-link-aliased  to the access path
corresponding to  the lhs  of statement $i$  at $\In{i}$. \Lhs($i$)  denotes the
access  path  corresponding  to  the  lhs access  expression  of  assignment  in
statement $i$.  $pred(i)$ is  the set  of predecessors of  statement $i$  in the
program.  $\mbox{\em ProperPrefix}(\rho)$  is the  set  of  all  proper prefixes  of
$\rho$.

We insert assignment \mbox{$\alpha = \NULL$} at program point $p$ if
$\nullify(\rho, p)$ is true. 

\subsection{Computing Candidate Access Paths for \NULL\ Insertion}
\label{sec:null.candidates} 
The method described above only checks whether a given
access path $\rho$ can be nullified at a given program point $p$.
We can generate the {\em candidate} set of access paths for
\NULL\ insertion at $p$ as follows: For any candidate access path
$\rho$, $\Base(\rho)$ must either be available or anticipable at
$p$. Additionally, all simple access paths are also candidates for
\NULL\ insertions. Therefore,
\begin{eqnarray}
  \cand(p) &=& \left\{ \rho\myarrow f \mid \rho \in \RCH{p},
  f \in \outfield(\rho)\right\} \nonumber\\
  && \cup \left\{\rho \mid \rho \mbox{ is a simple access path }
  \right\}\label{eq:candidate}
\end{eqnarray}
Where $\outfield(\rho)$ is the set of fields which can be used to
extend access path $\rho$ at $p$. It can be obtained easily from the
type information of the object $\Target(\rho)$ at $p$.

\begin{figure}[p]
\begin{center}
  \includegraphics{fig-null-insertion.epsi}
\end{center}

\caption{Null insertion for the program in
  Figure~\ref{fig:memory.graph}.}
\label{fig:null-insertion}
\rule{\textwidth}{.2mm}

\end{figure}

Note that all the information required for
equations~(\ref{eq:cannullify}), (\ref{eq:nullify:A}),
(\ref{eq:nullify:B}), and~(\ref{eq:candidate}) is obtained from the result
of data flow analyses described in preceding sections. Type
information of objects required by equation~(\ref{eq:candidate}) can be
obtained from the front end of compiler.
\Transp\ uses  may alias information  as computed in  terms of pairs  of access
graph.

\begin{example}
\label{exmp:null-insertion}
Figure~\ref{fig:null-insertion} lists a trace of the null insertion
algorithm for the program in Figure~\ref{fig:memory.graph}.
%
\mybox
\end{example}

\subsection{Reducing Redundant \NULL\ Insertions}
Consider a program with an assignment statement \mbox{$i: \alpha_x =
\alpha_y$}.  Assume a situation where, for some non-empty suffix
$\sigma$, both \mbox{$\nullify(\rho_y\myarrow\sigma, \In{i})$} and
\mbox{$\nullify(\rho_x\myarrow\sigma,\Out{i})$} are true. In that
case, we will be inserting $\alpha_y.\sigma = \NULL$ at $\In{i}$ and
$\alpha_x.\sigma = \NULL$ at $\Out{i}$. Clearly, the latter \NULL\
assignment is redundant in this case and can be avoided by checking if
$\rho_y\myarrow\sigma$ is nullable at $\In{i}$.

If must-alias analysis is performed then redundant assignments can be
reduced further. Since
must-link-alias relation is symmetric, reflexive, and transitive and
hence an equivalence relation, the set of candidate paths
at a program point can be divided into equivalence classes based on
must-link-alias relation. Redundant \NULL\ assignments can be reduced
by nullifying at most one access path in any equivalence class.

\section{Convergence of Heap Reference Analysis}
\label{sec:termination}

The \NULL\ assignment insertion algorithm makes a single traversal
over the control flow graph. We show the termination of 
liveness analysis using the properties of access graph
operations. Termination of availability and anticipability can be
shown by similar arguments over finite sets of bounded access paths.
Termination of alias analysis follows from \citeN{hind99interprocedural}.

\subsection{Monotonicity}
\label{sec:access.graph.properties}

For a program there are a finite number of basic blocks, a finite
number of fields for any root variable, and a finite number of field
names in any access expression. Hence the number of access graphs for
a program is finite. Further, the number of nodes and hence the size
of each access graph, is bounded by the number of labels which can be
created for a program.

Access graphs for a variable $x$ form a complete lattice with a
partial order $\sqsubseteq_G$ induced by $\cupG$.  Note that $\cupG$
is commutative, idempotent, and associative.
Let \mbox{$G = \langle x,N_F,N_I,E\rangle$} and \mbox{$G' = \langle
x,N'_F,N'_I,E'\rangle$} where subscripts $F$ and $I$ distinguish
between the final and intermediate nodes. The partial 
order $\sqsubseteq_G$ is defined as
\begin{equation*}
  G \sqsubseteq_G G' \Leftrightarrow 
    \left(N'_F \subseteq N_F\right) \wedge 
    \left(N'_I \subseteq \left(N_F \cup N_I\right)\right) \wedge
    \left(E' \subseteq E\right)
\end{equation*}
Clearly, $G \sqsubseteq_G G'$ implies that $G$ contains all access
paths of $G'$.  We extend $\sqsubseteq_G $ to a set of access graphs
as follows:
\[
S_1 \sqsubseteq_{S} S_2 \Leftrightarrow
   \forall G_2 \in S_2, \exists G_1 \in S_1
   \mbox{ s.t. } G_1 \sqsubseteq_G G_2
\]
It is easy to verify that $\sqsubseteq_G$ is reflexive, transitive,
and antisymmetric.  For a given variable $x$, the access graph
$\Empty\!_G$ forms the $\top$ element of the lattice while the $\bot$
element is a greatest lower bound of all access graphs.

The partial order over access graphs and their sets can be carried
over unaltered to remainder graphs ($\sqsubseteq_{RG}$) and their
sets ($\sqsubseteq_{RS}$), with the added condition that $\EFG$
is incomparable to any other non empty remainder graph. 

\begin{figure}[t]
\begin{tabular}{|l|r@{\ \ }c@{\ \ }l|}
\hline
Operation & \multicolumn{3}{|c|}{Monotonicity} \\ \hline \hline
Union &  $G_1\sqsubseteq_G G_1' \wedge G_2\sqsubseteq_G G_2'$
	& $\Rightarrow $ 
	& $G_1 \cupG G_2 \sqsubseteq_G G_1' \cupG G_2'$
          \rule[-.15cm]{0cm}{.45cm}
	\\ \hline
Path Removal & $G_1\sqsubseteq_G G_2 $
        &$\Rightarrow $
	&$G_1\minus\rho \sqsubseteq_G G_2 \minus\rho$\rule[-.15cm]{0cm}{.45cm} 
	\\ \hline
Factorization & $G_1\sqsubseteq_G G_2 $
        &$\Rightarrow $
	&$G_1/(G,M) \sqsubseteq_{RS} G_2/(G,M)$\rule[-.15cm]{0cm}{.45cm} 
	\\ \hline
Extension & $RS_1 \sqsubseteq_{RS} RS_2 \wedge G_1 \sqsubseteq_G G_2  \wedge
	M_1 \subseteq M_2$
        &$\Rightarrow$
	&$\extend{(G_1,M_1)}{RS_1} \sqsubseteq_{G} \extend{(G_2,M_2)}{RS_2} $
           \rule[-.15cm]{0cm}{.45cm} 
	\\ \hline
\renewcommand{\arraystretch}{.8}%
\begin{tabular}{@{}l@{}}
Link-Alias\\
Closure
\end{tabular}
        & $G_1\sqsubseteq_G G'_1 \wedge G_2\sqsubseteq_G G'_2 $
	&$\Rightarrow$
	&$\LnG(G_1,G_2,\langle g_x,g_y\rangle)\sqsubseteq_S\LnG(G'_1,G'_2,\langle g_x,g_y\rangle)$
         \rule[-.15cm]{0cm}{.6cm}
	\\ \hline
\end{tabular}
\caption{Monotonicity of Access Graph Operations}
\rule{\textwidth}{.2mm}
\label{fig:monotonicity.ag}.
\end{figure}

Access graph operations are monotonic as described in
Figure~\ref{fig:monotonicity.ag}.  Path removal is monotonic
in the first argument but not in the second argument. Similarly factorization is
monotonic in the first argument but not in the second and the third
argument. However, we show that in each context where they are used,
the resulting functions are monotonic:
\begin{enumerate}
\item Path removal is used only for an assignment
  \mbox{$\alpha_x=\alpha_y$}.  It is used in liveness analysis and its second argument 
  is $\rho_x$ which is constant for
  any assignment statement \mbox{$\alpha_x=\alpha_y$}. Thus the resulting
  flow functions are monotonic.
\item Factorization is used in the following situations:
  \begin{enumerate}
  \item {\em Link-alias closure of access graphs}.  
	From equation~(\ref{eq:link.alias.computation}) it is clear \LnG\ is
	monotonic in the first argument (because it is used in $\cupG$) and
	the second argument (because it is supplied as the first argument of
	factorization). The third and the fourth arguments of \LnG\ are linear 
	access graphs containing a single path and hence are incomparable with
	any other linear access graph.
    Thus link-alias computation is monotonic in all its arguments.

  \item {\em Liveness analysis}.  Factorization is used for the flow
    function corresponding to an assignment \mbox{$\alpha_x=\alpha_y$}
    and its second argument is $\graphA{\rho_x}$ while its third
    argument is $\lNode{\graphA{\rho_x}}$ both of which are
    constant for any assignment statement
    \mbox{$\alpha_x=\alpha_y$}. Thus, the resulting flow functions are
    monotonic.
  \end{enumerate}
\end{enumerate}
Thus we conclude that all flow functions are monotonic.  Since
lattices are finite, termination of heap reference analysis follows.

{
Appendix~\ref{sec:non-distributivity} discusses the distributivity of 
flow functions.
}

\subsection{Complexity}
\label{sec:complexity}

This section discusses the issues which influence the complexity and
efficiency of performing heap reference analysis. Empirical
measurements which corroborate the observations made in this section
are presented in Section~\ref{sec:measurements}.

The data flow frameworks defined in this paper are not {\em
separable\/}~\cite{dfa.chap} because the data flow information of a
variable depends on the data flow information of other variables.
Thus the number of iterations over control flow graph is not bounded
by the depth of the graph~\cite{asu,hect,dfa.chap} but would also
depend on the number of root variables which depend on each other.

Although we consider each statement to be a basic block, our control
flow graphs retain only statements involving references. A further
reduction in the size of control flow graphs follows from the fact
that successive use statements need not be kept separate and can be
grouped together into a block which ends on a reference assignment.

The amount of work done in each iteration is not fixed but depends on
the size of access graphs. Of all operations performed in an
iteration, only $\states{G}{G'}$ is costly. Conversion to
deterministic access graphs is also a costly operations but is
performed for a single pass during \NULL\ assignment insertion.  In
practice, the access graphs are quite small because of the following
reason: Recall that edges in access graphs capture dependence of a
reference made at one program point on some other reference made at
another point (Section~\ref{sec:dependence}). 
In real  programs, traversals  involving long dependences  are performed
using  iterative constructs  in the  program.  In  such  situations, the
length  of  the  chain of  dependences  is  limited  by the  process  of
summarization because summarization treats nodes with the same
label  as  being identical.  Thus,  in  real  programs chains  of  such
dependences, and hence the access graphs, are quite small in size.
 This is corroborated by
Figure~\ref{tab:space.time.data} which provides the empirical data for
the access graphs in our examples.  The average number of nodes in
these access graphs is less than 7 while the average number of edges
is less than 12.  These numbers are still smaller in the interprocedural
analysis. Hence the complexities of access graph operations is
not a matter of concern.

\section{Safety of \NULL\ Assignment Insertion}
\label{sec:soundness}

We have to prove that the \NULL\ assignments inserted by our algorithm
(Section~\ref{sec:nullability}) in a program are safe in that they do
not alter the result of executing the program. We do this by showing
that (a) an inserted statement itself does not raise a dereferencing exception, and
(b) an inserted statement does not affect any other statement, both
original and inserted.

We use the subscripts $b$ and $a$ for a program point~$p$ to denote
``before'' and ``after'' in an execution order. Further, the
corresponding program points in the original and modified program are
distinguished by the superscript $o$ and $m$. The correspondence is
defined as follows: If \Pmod\ is immediately before or after an
inserted assignment \mbox{$\alpha = \NULL$}, \Porg\ is the point where
the decision to insert the \NULL\ assignment is taken.  For any other
\Pmod, there is an obvious \Porg.

We first assert the soundness of availability, anticipability and
alias analyses without proving them.

\begin{lemma}
\label{lemm:avail.sound}
{\rm (Soundness of Availability Analysis).} Let $AV_{\!\!\Pa}$ be the
set of access paths available at program point \Pa. Let $\rho \in
AV_{\!\!\Pa}$.  Then along every path reaching \Pa, there exists a
program point \Pb, such that {the link represented by} $\Front(\rho)$
is either dereferenced or assigned a non-\NULL\ l-value at \Pb\ and is
not made \NULL\ between \Pb\ and \Pa.
\end{lemma}

\begin{lemma}
\label{lemm:ant.sound}
{\rm (Soundness of Anticipability Analysis).} Let $AN_p$ be the set of
access paths anticipable at program point~$p$. Let $\rho \in
AN_p$. Then along every path starting from $p$, {the link represented
by $\Front(\rho)$} is dereferenced before being assigned.
\end{lemma}

For  semantically valid input  programs (i.e.\  programs which  do not
generate               dereferencing              exceptions),
Lemma~\ref{lemm:avail.sound}  and Lemma~\ref{lemm:ant.sound} guarantee
that {if}  $\rho$ is available or anticipable  at $p$, $\Target(\rho)$
can be dereferenced at $p$.

\begin{lemma}
\label{lemm:soundness.aliasing} 
{\rm (Soundness of Alias Analysis).} Let $\Front(\rho_x)$ represents
the same link as $\Front(\rho_y)$ at a program point~$p$ during some
execution of the program. Then link-alias computation of $\rho_x$ at
$p$ would discover $\rho_y$ to be link-aliased to $\rho_x$.
\end{lemma}

For the main claim, we relate the access paths at \Pa\ to the
access paths at \Pb\ by incorporating the effect of intervening
statements only, regardless of the statements executed before \Pb.  In
some execution of a program, let $\rho$ be the access path of interest
at \Pa\ and the sequence of statements between \Pb\ and \Pa\ be $s$.\footnote{
  When $s$ is a function call $\alpha_x = f(\alpha_y)$, \Pa\ is the entry point
  of $f$ and \Pb\ is the program point just before the statement $s$ in the
  caller's body. Analogous remark holds for the return statement.}
Then \mbox{$T(s,\rho)$} represents the access path at \Pb\ which, if
non-\Empty, can be used to access the link represented by
$\Front(\rho)$.   \mbox{$T(s, \rho)$} captures the
transitive effect of backward transfers of $\rho$ through $s$.  $T$ is
defined as follows:
\begin{eqnarray*}
T(s, \rho ) & = & 
\left\{\begin{array}{ll}
	\rho & s \mbox{ is a use statement } \\
	\rho & s \mbox{ is } \alpha_x = \ldots \mbox{ and }
		\rho_x \mbox{ is not a prefix of } \rho \\
	\Empty & s \mbox{ is } \alpha_x = \mbox{{\em New\/} and } 
		\rho=\rho_x\myarrow\sigma \\
	\Empty & s \mbox{ is } \alpha_x = \NULL\ \mbox{ and }
		\rho=\rho_x\myarrow\sigma \\
	\rho_y\myarrow\sigma & s \mbox{ is } \alpha_x = \alpha_y \mbox{ and }
		\rho=\rho_x\myarrow\sigma \\ 
    \rho & s \mbox{ is the function call } \alpha_x = f(\alpha_y) \mbox{ and }
	\\ & \RootVar(\rho) \mbox{ is a global variable}\\
    \rho_y\myarrow\sigma & s \mbox{ is the function call } \alpha_x = f(\alpha_y),
	\rho = z\myarrow\sigma \mbox{ and }\\ & z \mbox{ is the formal parameter of } f\\
    \rho & s \mbox{ is the return statement } return(\alpha_z) \mbox{ and }
	\\ & \RootVar(\rho) \mbox{ is a global variable}\\
    \rho_z\myarrow\sigma & s \mbox{ is the return statement } 
    return(\alpha_z), \rho = \rho_x\myarrow\sigma \mbox{ and}\\
         & \mbox{the corresponding  call is } \alpha_x = f(\alpha_y)\\

	T(s_1, T(s_2,\rho)) & s \mbox{ is  a sequence } s_1;s_2
\end{array}
\right. 
\end{eqnarray*}

\begin{lemma}{\em (Liveness Propagation)}.
\label{lemm:liveness.prop}
Let $\rho^a$ be in some explicit liveness graph at \Pa. Let the
sequence of statements between \Pb\ to \Pa\ be $s$. Then, if
\mbox{$T(s,\rho^a) = \rho^b$} and $\rho^b$ is not \Empty, then
$\rho^b$ is in some explicit liveness graph at $p_b$.
\end{lemma}
\begin{proof}
The proof is by structural induction on $s$.
Since $\rho^b$ is
non-\Empty, the base cases are:
\begin{enumerate}
\item $s$ is a use statement. In this case $\rho^b = \rho^a$. \label{base.step.1}
\item $s$ is an assignment \mbox{$\alpha_x = \ldots$} such that
  $\rho_x$ is not a prefix of $\rho^a$. Here also $\rho^b = \rho^a$.
  \label{base.step.2}
\item $s$ is an assignment \mbox{$\alpha_x = \alpha_y$} such that
  \mbox{$\rho^a = \rho_x\myarrow\sigma$}. In this case $\rho^b = \rho_y\myarrow\sigma$.
  \label{base.step.3}
\item \label{base.step.4} $s$ is the function call $\alpha_x = f(\alpha_y)$. The
  only interesting  case is  when $\rho^a =  z\myarrow\sigma$, where $z$  is the
  formal parameter of $f$. In this case, $\rho^b = \rho_y\myarrow\sigma$.
\item \label{base.step.5}  $s$ is  the return statement  $return(\alpha_z)$. The
only  interesting  case  is   when  $\rho^a  =  \rho_x\myarrow\sigma$,  and  the
corresponding  call  is  $\alpha_x  =   f(\alpha_y)$.  In  this  case,  $\rho^b  =
\rho_z\myarrow\sigma$.
\end{enumerate}
For (\ref{base.step.1}) and (\ref{base.step.2}), since
$\rho^a$ is not in \Lkill{}, $\rho^b$ is in some explicit liveness
graph at \Pb. For (\ref{base.step.3}), 
from Equation~(\ref{eq:xfer.y.asgn}), $\rho^b$ is in some explicit liveness 
graph at \Pb.
 For (\ref{base.step.4}) and (\ref{base.step.5}), the result
follows from the fact that $\Summary(\rho_y)$ and $\Summary(\rho_z)$ are
in the explicit liveness graph of the program points before the call and return
statements respectively.

For the inductive step, assume that the lemma holds for $s_1$ and
$s_2$.  From the definition of $T$, there exists a non-\Empty\
$\rho^i$ at the intermediate point~$p_i$ between $s_1$ and $s_2$, such
that \mbox{$\rho^i = T(s_2,\rho^a)$} and \mbox{$\rho^b =
T(s_1,\rho^i)$}. Since $\rho^a$ is in some explicit liveness graph at
\Pa, by the induction hypothesis, $\rho^i$ must be in some explicit
liveness graph at $p_i$.  Further, by the induction hypothesis,
$\rho^b$ must be in some explicit liveness graph at \Pb.
\end{proof}
\begin{lemma}
\label{lemm:corresponding.liveness}
Every access path which is in some  liveness graph at \Pbmod\
is also in some liveness graph at \Pborg.
\end{lemma}
\begin{proof}
If an extra explicitly live access path is  introduced in the
modified program, it could be only because of an inserted assignment
\mbox{$\alpha = \NULL$} at some \Pamod.  The only access paths which
this statement can add to an explicit liveness graph are the paths
corresponding the proper prefixes of $\alpha$. However, the algorithm
selects $\alpha$ for nullification only if the access paths
corresponding to all its proper prefixes are in some explicit liveness
graph.  Therefore every access path which is in some explicit liveness
graph at \Pamod\ is also in some explicit liveness graph at
\Paorg. The same relation would hold at \Pbmod\ and \Pborg.

 If an extra live access path is introduced in the modified program, it
could be only because of an  inserted assignment \mbox{$\alpha = \NULL$} at some
\Pamod.   The only  access paths  which this  statement can  add to  an liveness
graphs are  $\LnA(\rho', \Aset^m)$, where $\rho'$  is a proper  prefix of $\rho$
and  $\Aset^m$ represents  the  alias  set at  \Pamod.   However, the  algorithm
selects  $\alpha$  for  nullification  at  \Pamod\  only  if  the  access  paths
corresponding  to  all  its  proper  prefixes  are in  some  liveness  graph  at
\Paorg. As liveness graphs are closed under link aliasing, this implies that the
liveness graph at \Paorg\ includes paths $\LnA(\rho', \Aset^o)$, where $\Aset^o$
represents the  alias set  at \Paorg.  Since  inserted statements can  only kill
aliases, $\Aset^m  \subseteq \Aset^o$.  Thus, $\LnA(\rho',  \Aset^m)$, the paths
resulting out of insertion, are also in the liveness graph at \Paorg.  Therefore
every access  path which is in  some liveness graph  at \Pamod\ is also  in some
liveness graph at \Paorg.  The same relation would hold at \Pbmod\ and \Pborg.
%
\end{proof}
\begin{theorem}{\em  (Safety of \NULL\ insertion)}.  Let the  assignment
\mbox{$\alpha^b = \NULL$} be inserted by the algorithm immediately before
\Pbmod. Then:
\begin{enumerate}
\item Execution of \mbox{$\alpha^b = \NULL$} does not raise any
  exception due to dereferencing.
\item Let $\alpha^a$ be used immediately after \Pamod\ (in an original
  statement or an inserted \NULL\ assignment).  Then, execution of
  \mbox{$\alpha^b = \NULL$} cannot nullify any link used in
  $\alpha^a$.
\end{enumerate}
\end{theorem}
\begin{proof} We prove the two parts separately. 
\begin{enumerate}
\item If $\alpha^b$ is a root variable, then the execution of
  \mbox{$\alpha^b = \NULL$} cannot raise an exception. When $\alpha^b$
  is not a root variable, from the null assignment algorithm, every
  proper prefix $\rho'$ of $\rho^b$ is either anticipable or
  available. From the soundness of both these analyses,
  $\Target(\rho')$ exists and the execution of \mbox{$\alpha^b =
    \NULL$} cannot raise an exception.
\item We prove this by contradiction.  Let $s$ denote the sequence of statements
  between \Pbmod\ and \Pamod.  Assume that \mbox{$\alpha^b = \NULL$} nullifies a
  link  used in  $\alpha^a$.  This  is possible  only if  there exists  a prefix
  $\rho'$ of $\rho^a$  such that $T(s,\rho')$ shares its  frontier with $\rho^b$
  at  \Pbmod.
 By  Lemma~\ref{lemm:liveness.prop}, $T(s,\rho')$  must be  in some
  explicit liveness  graph at \Pbmod.   From Lemma~\ref{lemm:soundness.aliasing}
  and  the  definition  of liveness,  $\rho^b$  is  in  some liveness  graph  at
  \Pbmod. By  Lemma~\ref{lemm:corresponding.liveness}, $\rho^b$ is  also in some
  liveness graph at \Pborg. Thus  a decision to insert \mbox{$\alpha^b = \NULL$}
  cannot be taken at \Pborg.
\end{enumerate}
\end{proof}

\section{Empirical Measurements}
\label{sec:measurements}

\begin{figure}[t]
\begin{tabular}{c|c}
  \includegraphics{Loop.epsi} &
  \includegraphics{DLoop.epsi} \\ \hline & \\
  \includegraphics{CReverse.epsi} &
  \includegraphics{BiSort.epsi} \\\hline & \\
  \includegraphics{TreeAdd.epsi} &
  \includegraphics{GCBench.epsi} \\ \hline
  \multicolumn{2}{c}{}
\end{tabular}

X axis indicates measurement instants in milliseconds.  Y axis
indicates heap usage in KB. Solid line represents memory required for
original program while dashed line represents memory for the modified
program. Observe that the modified program executed faster than the
original program in each case.
\caption{Temporal plots of memory usages.}  
\label{fig:plots}
\rule{\textwidth}{.2mm}
\end{figure}

In order to show the effectiveness of heap reference analysis, we have
developed proof-of-concept implementations
of heap reference analysis at two levels:
One at the interprocedural level and the other at the intraprocedural level.

\subsection{Experimentation Methodology}

Our intraprocedural analyzer, which predates the interprocedural version
is an evidence of the effectiveness  of  intraprocedural  analysis.
It was implemented using 
XSB-Prolog\footnote{Available from \url{http://xsb.sourceforge.net}.}.
The measurements were made
on a 800 MHz Pentium III machine with 128 MB memory running Fedora
Core release 2. The benchmarks used were \texttt{Loop},
\texttt{DLoop}, \texttt{CReverse}, \texttt{BiSort}, \texttt{TreeAdd}
and \texttt{GCBench}.  Three of these (\texttt{Loop}, \texttt{DLoop}
and \texttt{CReverse}) are similar to those
in~\cite{ran.shaham-sas03}.  \texttt{Loop} creates a singly linked
list and traverses it, \texttt{DLoop} is doubly linked list variation
of the same program, \texttt{CReverse} reverses a singly linked
list. \texttt{BiSort} and \texttt{TreeAdd} are taken from Java version
of Olden benchmark suite~\cite{jolden}.  \texttt{GCBench} is taken
from~\cite{gcbench}. 

For measurements on this implementation, the function of interest in a
given Java  program was manually translated  to Prolog representation.
This allowed us to avoid redundant information like temporaries, empty
statements etc. resulting in a compact representations of programs.
The interprocedural information for  this function was approximated in
the  Prolog   representations  in  the  following   manner:  Calls  to
non-recursive functions were inlined  and calls to recursive functions
were replaced by iterative  constructs which approximated the liveness
property of heap manipulations in  the function bodies.  The result of
the analysis  was used  to manually insert  \NULL\ assignments  in the
original Java programs to create modified Java programs.

Manual interventions allowed us to handle procedure calls
without performing interprocedural analysis. In order to automate the
analysis and extend it to interprocedural level, we  
used SOOT~\cite{vall99soot} which has built in support for
many of our requirements. However,
compared to the Prolog representation of programs,
the default Jimple representation used by SOOT is 
not efficient for our purposes because it introduces a large number of
temporaries and contains all statements even if they do not affect
heap reference analysis.

As was described earlier, our interprocedural analysis is very simplistic.
Our experience shows that imprecision of interprocedural
alias analysis increases the size of alias information
thereby making the analysis inefficient apart from reducing the precision 
of the resulting information. 
This effect has been worsened by the fact that SOOT introduces 
a large number of temporary variables. 
Besides, the complete alias information is not required for our purposes.

We believe that our approach can be made much more scalable by 
\begin{itemize}
\item Devising a method of avoiding full alias analysis and computing only 
       the required alias information, and
\item Improving the Jimple representation by eliminating redundant 
       information, combining multiple successive uses into a single
       statement etc.
\end{itemize} 

The implementations, along with the test programs (with their
original, modified, and Prolog versions) are available
at~\cite{hra.prototype}.

\newcommand{\NIter}{\mbox{\#Iter}}
\newcommand{\NGph}{\mbox{\#G}}
\newcommand{\Nnull}{\mbox{\#\NULL}}

\begin{figure}[t]
\begin{center}
\renewcommand{\arraystretch}{1}
\renewcommand{\rotatebox}[2]{#2}
\begin{tabular}{|@{}c@{}|}
\hline
Intraprocedural analysis of selected method (Prolog Implementation) \\ \hline \hline
\begin{tabular}{@{\ }l@{\ }|r|r|r|r|r|r|r|r|r|r}
Program
& \multicolumn{2}{|c|}{Analysis} 
& \multicolumn{5}{|c|}{Access Graphs}
& \multicolumn{1}{|c|}{}
& \multicolumn{2}{|c@{}}{Execution} 
\\ \cline{2-3}\cline{4-8}
Name.Function
& \multicolumn{1}{|c|}{} 
& \multicolumn{1}{|c|}{Time} 
& \multicolumn{1}{|c|}{}
& \multicolumn{2}{|c|}{Nodes}
& \multicolumn{2}{|c|}{Edges}
& \multicolumn{1}{|c|}{}
& \multicolumn{2}{|c@{}}{Time (sec)} 
\\ \cline{5-8}\cline{10-11}
& \multicolumn{1}{|c|}{\rotatebox{90}{\NIter}}
& \multicolumn{1}{|c|}{(sec)} 
& \multicolumn{1}{|c|}{\NGph}
& \multicolumn{1}{|c|}{\rotatebox{90}{Avg}}
& \multicolumn{1}{|c|}{\rotatebox{90}{Max}}
& \multicolumn{1}{|c|}{\rotatebox{90}{Avg}}
& \multicolumn{1}{|c|}{\rotatebox{90}{Max}}
& \multicolumn{1}{|c|}{\rotatebox{90}{\Nnull}}
& \multicolumn{1}{|c|}{Orig.} 
& \multicolumn{1}{|c@{}}{Mod.}
\\
\hline\hline
\texttt{Loop.main}    & 5& 0.082 & 172 & 1.13 & 2 &  0.78 &  2 &  9
& 1.503 & 1.388 \\ \hline
\texttt{DLoop.main}   & 5& 1.290 & 332 & 2.74 & 4 &  5.80 & 10 & 11
& 1.594 & 1.470 \\ \hline
\texttt{CReverse.main}& 5& 0.199 & 242 & 1.41 & 4 &  1.10 &  6 &  8
& 1.512 & 1.414 \\ \hline
\texttt{BiSort.main}  & 6& 0.083 &  63 & 2.16 & 3 &  3.81 &  6 &  5
& 3.664 & 3.646 \\ \hline
\texttt{TreeAdd.addtree} & 6& 0.255 & 132 & 2.84 & 7 &  4.87 & 14 &  7
& 1.976 & 1.772 \\ \hline
\texttt{GCBench.Populate} & 6& 0.247 & 136 & 2.73 & 7 &  4.63 & 14 &  7
& 132.99 & 88.86 
\end{tabular}
\\
\hline \hline
Interprocedural analysis of all methods (SOOT Implementation) \\ \hline
\hline
\begin{tabular}{l|r|r|r|r|r|r|r|r|r|r}
\multicolumn{1}{c|}{Program}
& \multicolumn{1}{|c|}{LOC}
& \multicolumn{1}{|c|}{\#}
& \multicolumn{1}{|c|}{Analysis}
& \multicolumn{4}{|c|}{Access Graph Stats}
& \multicolumn{1}{|c|}{} 
& \multicolumn{2}{|c}{Execution} 
\\ \cline{5-8}
\multicolumn{1}{c|}{Name}
& \multicolumn{1}{|c|}{in}
& \multicolumn{1}{|c|}{methods}
& \multicolumn{1}{|c|}{Time}
& \multicolumn{2}{|c|}{Nodes}
& \multicolumn{2}{|c|}{Edges}
& \multicolumn{1}{|c|}{\Nnull} 
& \multicolumn{2}{|c}{Time (sec)} 
\\ \cline{5-8}\cline{10-11}
\multicolumn{1}{c|}{}
& \multicolumn{1}{|c|}{Jimple}
& \multicolumn{1}{|c|}{}
& \multicolumn{1}{|c|}{(sec.)}
& \multicolumn{1}{|c|}{Max}
& \multicolumn{1}{|c|}{Avg}
& \multicolumn{1}{|c|}{Max}
& \multicolumn{1}{|c|}{Avg}
& \multicolumn{1}{|c|}{} 
& \multicolumn{1}{|c|}{Orig.} 
& \multicolumn{1}{|c}{Mod.}
\\ \hline\hline
\texttt{Loop}    &  83 &  2 &  0.558  & 2 & 1.24 & 2 & 0.39 & 12 & 1.868 & 1.711 \\ \hline
\texttt{DLoop}   &  78 &  2 & 20.660  & 5 & 1.45 & 12 & 0.76 & 12 & 1.898 & 1.772\\ \hline
\texttt{CReverse}&  85 &  2 &  1.833  & 3 & 1.39 & 4 & 0.51 & 12 & 1.930 & 1.929 \\ \hline
\texttt{BiSort}  & 466 & 12 &  1.498  & 7 & 1.29 & 10 & 0.40 & 77 & 1.519 & 1.524 \\ \hline
\texttt{TreeAdd} & 228 &  4 &  0.797  & 6 & 1.29 & 7 & 0.46 & 34 & 2.704 & 2.716\\ \hline
\texttt{GCBench} & 226 &  9 &  1.447  & 4 & 1.13 & 5 & 0.16 & 56 & 122.731 & 60.372 \\ 
\hline
\end{tabular}
\\ \multicolumn{1}{c}{}
\end{tabular}

\hspace*{.25in}
\begin{itemize}
\item[-] \NIter\ is the maximum number of iterations taken by any analysis. 
\item[-] Analysis Time is the total time taken by all analyses. 
\item[-] \NGph\ is total number of access graphs created by alias
analysis and liveness analysis. Prolog implementation performs alias analysis also
using access graphs.
\item[-] Max nodes (edges) is the maximum over number of nodes (edges) in
all access graphs.
In some cases, maximum number nodes/edges is more in case of intraprocedural 
analysis due to presence of longer paths in explicitly
supplied boundary information, which gets replaced by a single *
node in interprocedural analysis.
\item[-] Avg nodes (edges) is the average number of nodes (edges) over
all access graphs.
\item[-] \Nnull\ is the number of inserted \NULL\ assignments.
\end{itemize}
\end{center}
\vspace*{-.15in}
\caption{Empirical measurements of proof-of-concept implementations of heap reference
analyzer.}
\label{tab:space.time.data}
\rule{\textwidth}{.2mm}
\end{figure}

\subsection{Measurements and Observations}
\label{sec:measurements.obs}

Our experiments were directed at measuring:
\begin{enumerate}
\item {\em The efficiency of analysis}.  We measured the total time
  required, number of iterations of round robin analyses, and the
  number and sizes of access graphs.
\item {\em The effectiveness of \NULL\ assignment insertions}.  The
  programs were made to create huge data structures.  Memory usage was
  measured by explicit calls to garbage collector in both modified and
  original Java programs at specific probing points such as call
  sites, call returns, loop begins and loop ends.  The overall
  execution time for the original and the modified programs was also
  measured.
\end{enumerate}

The results of our experiments are shown in Figure~\ref{fig:plots} and
Figure~\ref{tab:space.time.data}. As can be seen from
Figure~\ref{fig:plots}, nullification of links helped the garbage
collector to collect a lot more garbage, thereby reducing the
allocated heap memory. In case of BiSort, however, the links were last
used within a recursive procedure which was called multiple times.
Hence, safety criteria prevented \NULL\ assignment insertion within
the called procedure. Our analysis could only nullify the root of the
data structure at the end of the program. Thus the memory was released
only at the end of the program.

For interprocedural analysis, class files for both original as well as
modified  programs were  generated using  SOOT.  As can  be seen  from
Figure~\ref{tab:space.time.data},     modified    programs    executed
faster. In general, a reduction in execution time can be attributed to
the following  two factors: (a) a  decrease in the number  of calls to
garbage  collector and  (b) reduction  in the  time taken  for garbage
collection  in   each  call.  The   former  is  possible   because  of
availability of a larger amount of free memory, the latter is possible
because  lesser  reachable memory  needs  to be  copied.\footnote{This
happens  because   Java  Virtual   Machine  uses  a   copying  garbage
collector.} In  our experiments,  factor~(a) above was  absent because
the  number  of  (explicit)  calls  to  garbage  collector  were  kept
same. \texttt{GCBench}  showed a  large improvement in  execution time
after  \NULL\ assignment insertion.  This is  because \texttt{GCBench}
creates large trees  in heap, which are not used  in the program after
creation and  our implementation  was able to  nullify left  and right
subtrees of  these trees immediately  after their creation.  This also
reduced the high water mark of the heap memory requirement.

As explained in Section~\ref{sec:complexity}, sizes of the access
graphs (average number of nodes and edges) is small. This can be
verified from Figure~\ref{tab:space.time.data}. The analysis of
\texttt{DLoop} creates a large number of access graphs because of the
presence of cycles in heap. In such a case, a large number of alias
pairs are generated, many of which are redundant. Though it is
possible to reduce analysis time by eliminating redundant alias pairs,
our implementation, being a proof-of-concept implementation, does not do so 
for sake of simplicity.

Our technique and implementation compares well with the
technique and results described in~\cite{ran.shaham-sas03}.  
A conceptual comparison with this method is included in Section~\ref{sec:ran.comparison}. 
The implementation described in~\cite{ran.shaham-sas03} runs on a 900 MHz
P-III with 512 MB RAM running Windows 2000. It takes 1.76 seconds,
2.68 seconds and 4.79 seconds respectively for \texttt{Loop},
\texttt{DLoop} and \texttt{CReverse} for \NULL\ assignment insertion.
Time required by our implementation for the above mentioned
programs is given in Figure~\ref{tab:space.time.data}. Our
implementation automatically computes the program points for \NULL\
insertion whereas their method cannot do so.
{Our implementation} performs much
better { in all cases.} 

\section{Extensions for C++ }
\label{sec:c++ext}

This approach becomes applicable to C++ by extending the concept of access graphs
to faithfully represent the C++ memory model. It is assumed that the memory 
which becomes unreachable due to nullification of pointers is reclaimed by
an independent garbage collector. Otherwise, explicit reclamation of
memory can be performed by checking that no node-alias of a nullified
pointer is live.

In order to extend the concept of access graphs to C++, 
we need to account for two major differences between the C++ and
the Java memory model:
\begin{enumerate}
\item Unlike Java, C++ has explicit pointers.  Field of a structure
  ({\tt struct} or {\tt class}) can be accessed in two different ways
  in C++: 
  \begin{itemize}
  \item using pointer dereferencing ({$*.$}),
  e.g. {$(*x).\lptr$}\footnote{This is equivalently written as
  $x\!-\!\!\!\!\!>\!\!\lptr$.} or 
  \item using simple dereferencing
  ({$.$}) , e.g. {$y.\rptr$}. 
  \end{itemize} 
  We need to distinguish between the two.
  
\item Although root variables are allocated on stack in both C++ and
  Java, C++ allows a pointer/reference to point to root variables on stack 
  through the use of addressof ({\tt\&}) operator, whereas
  Java does not allow a reference to point to stack. 
  Since the root nodes in access graphs  do not have an incoming edge by
  definition, it is not possible to use access graphs
  directly to represent memory links in C++.  
\end{enumerate}

\newcommand{\deref}{\mbox{\sf\em deref}}

We create access graphs for C++ memory model as follows:
\begin{enumerate}
\item We treat dereference of a pointer as a field reference, i.e., $*$
is considered as a field named \deref. For example, an access expression $(*x).\lptr$
is viewed as $ x.\deref.\lptr$, and corresponding access path is
\mbox{$x\myarrow\deref\myarrow\lptr$}. The access path for $x.\lptr$ is
\mbox{x\myarrow\lptr}. 

\item Though a pointer can point to a variable $x$, it is not
possible extract the address of $\&x$, i.e. no pointer can point to $\&x$. 
For Java, we partition memory as stack and
heap, and had root variables of access graphs correspond to stack
variables. In C++, we partition the memory as {\em address of variables}
and rest of the memory (stack and heap together). We make the roots of
access graphs correspond to addresses of variables. A root variable $y$
is represented as $\deref\,(\&y)$. Thus,
\scalebox{.9}{%
\rule[-.2cm]{0cm}{.6cm}%
\raisebox{1.35mm}{\rnode{n0}{}} \ \ \  \
\rule{0cm}{.4cm}\raisebox{.6mm}{\circlenode[framesep=.4mm]{n1}{\small$\&y$}}
\ \  \
\raisebox{.6mm}{\circlenode[framesep=.5mm]{n2}{$d_1$}}
\ncline{->}{n1}{n2}
\ \  \
\raisebox{.6mm}{\circlenode[framesep=.7mm]{n3}{$l_2$}}
\ncline{->}{n2}{n3}
\ncline[doubleline=true]{->}{n0}{n1}
}
represents access paths $\&y$ and \mbox{$\&y\myarrow \deref$} and
\mbox{$\&y\myarrow \deref\myarrow l$}, which correspond to access expressions
$\&y$, $y$ and $y.l$ respectively.

\end{enumerate}

Handling pointer arithmetic and type casting in C++ is orthogonal to
above discussion, and requires techniques similar
to~\cite{yong99pointer,cheng00modular} to be used.

\section{Related Work}
\label{sec:related}

Several properties of heap (viz. reachability, sharing, liveness etc.) have 
been explored in past; a good review has been provided by~\citeN{shape.chap}.  
In this section, we review the related work in the main
property of interest: liveness.  We are not aware of
past work in availability and anticipability analysis of heap
references.

\subsection{Liveness Analysis of Heap Data}
Most of the reported literature in liveness analysis of heap data
either does not address liveness of individual objects or
addresses liveness of objects identified by their allocation sites.  
Our method, by contrast, does not need the knowledge of allocation site.
Since the precision of a
garbage collector depends on its ability to distinguish between
reachable heap objects and live heap objects, even state of art
garbage collectors leave a significant amount of garbage
uncollected~\cite{AgesenDetMos98,shah00,shah01,shah02}.  All reported
attempts to incorporate liveness in garbage collection have been quite
approximate. The known approaches have been:
\begin{enumerate}
\item {\em Liveness of root variables.}  A popular approach (which has
  also been used in some state of art garbage collectors) involves
  identifying liveness of root variable on the stack. All heap objects
  reachable from the live root variables are considered
  live~\cite{AgesenDetMos98}.
\item {\em Imposing stack discipline on heap objects.}  These
  approaches try to change the statically unpredictable lifetimes of
  heap objects into predictable lifetimes similar to stack data. They
  can be further classified as
  \begin{itemize}
  \item {\em Allocating objects on call stack}.  These approach try to
    detect which objects can be allocated on stack frame so that they
    are automatically deallocated without the need of traditional
    garbage collection. A profile based approach which tracks the last
    use of an object is reported in~\cite{mcdo98}, while a static
    analysis based approach is reported in~\cite{reid99}.

    Some approaches ask a converse question: which objects are
    unstackable (i.e. their lifetimes outlive the procedure which
    created it)?  They use abstract interpretation and perform {\em
    escape analysis\/} to discover objects which {\em escape\/} a
    procedure\cite{Blanchet:1999:EAO,Blanchet:2003:EAJ,choi99escape}.
    All other objects are allocated on stack.
  \item {\em Associating objects with call stack}~\cite{cann00}. This
    approach identifies the stackability. The objects are allocated in
    the heap but are associated with a stack frame and the runtime
    support is modified to deallocate these (heap) objects when the
    associated stack frame is popped.
  \item {\em Allocating objects on separate stack}. This approach uses
    a static analysis called {\em region
    inference\/}~\cite{tofte98region,tofte-region-pldi-02} to identify
    {\em regions\/} which are storages for objects. These regions are
    allocated on a separate region stack.
  \end{itemize}
  All these approaches require modifying the runtime support for the
  programs.
\item {\em Liveness analysis of locally allocated objects.} The
  Free-Me approach~\cite{guyer06free}  combines  a   lightweight  pointer
analysis with liveness information that detects when allocated objects
die  and insert  statements to  free  such objects.  The analysis  is
simpler and  cheaper as the scope  is limited, but  it frees
locally  allocated   objects  only by separating objects which escape
the procedure call from those which do not.   The objects which do not
escape the procedure which creates them become unreachable at the end of the
procedure anyway and would be garbage collected. 
Thus their method merely advances the work
of garbage collection instead of creating new garbage. Further, this
does not happened in the called method. Further, their  method  uses
traditional liveness  analysis for root  variables only and  hence can
not free objects that are stored in field references. 
\item The {\em Shape Analysis Based\/} based approaches. The two approaches
     in this category are
     \begin{itemize}
      \item Heap Safety Automaton approach~\cite{ran.shaham-sas03} is a 
      recently reported work which
  comes closest to our approach since it tries to determine if a
  reference can be made \NULL.  We discuss this approach in the next
  section.
    \item \citeN{cherem06compile}   use  a   shape  analysis
framework~\cite{hackett05region} to  analyze a  single heap cell  to
discover  the  point in  the
program where it  object becomes unreachable. Their method  claims the objects
at such points  thereby reducing the work of  the garbage collector. They
use equivalence classes of expressions to store definite points-to and
definitely-not  points-to   information  in  order   to  increase  the
precision of abstract  reference counts.  However, multiple iterations
of the analysis and the optimization steps are required, since freeing
a cell  might result in  opportunities for more  deallocations.  Their
method  does not  take into  account the  last use  of an  object, and
therefore  does not make  additional objects  unreachable. 
     \end{itemize}
\end{enumerate}

\subsection{Heap Safety Automaton Based Approach}
\label{sec:ran.comparison}

This approach models safety of inserting a null statement at a given
point by an automaton. A shape graph based abstraction of the program
is then model-checked against the heap safety automaton.
Additionally, they also consider freeing the object; our approach can
be easily extended to include freeing.

The fundamental differences between the two approaches are
\begin{itemize}
\item Their method answers the following question: Given an access
  expression and a program point, can the access expression be set to
  \NULL \ immediately after that program point? However, they leave a
  very important question unanswered: Which access expressions should
  we consider and at which point in the program?  It is impractical to
  use their method to ask this question for every pair of access
  expression and program point.  Our method answers both the questions
  by finding out appropriate access expressions and program points.
\item We insert \NULL\ assignments at the earliest possible point.
  The effectiveness of any method to improve garbage collection
  depends crucially on this aspect.  Their method does not address
  this issue directly.
\item {As noted in Section~\ref{sec:measurements.obs},} their method is
  inefficient in practice. For a simple Java program containing 11
  lines of executable statements, it takes over 1.37 MB of storage and
  takes 1.76 seconds for answering the question: Can the variable $y$
  be set to \NULL \ after line 10?
\end{itemize}
Hence our approach is superior to their approach in terms of
completeness, effectiveness, and efficiency.

\section{Conclusions and Further Work }
\label{sec:conclusions}

Two fundamental challenges in analyzing heap data are that the
temporal and spatial structures of heap data seem arbitrary and are
unbounded.  The apparent arbitrariness arises due to the fact that the
mapping between access expressions and l-values varies dynamically.

The two key insights which allow us to overcome the above problems in the
context of liveness analysis of heap data are: 
\begin{itemize}
\item {\em Creating finite representations for properties of heap data using
  program structure}.
We create an abstract representation of heap in terms of sets of
access paths.  Further, a bounded representation, called access
graphs, is used for summarizing sets of access paths. Summarization is
based on the fact that the heap can be viewed as consisting of
repeating patterns which bear a close resemblance to the program
structure. Access graphs capture this fact directly by tagging program
points to access graph nodes.  Unlike
\cite{horw89dependence,ChaseWegZad90,choi93efficient,wilson95efficient,hind99interprocedural}
where only memory allocation points are remembered, we remember all
program points where references are used.  This allows us to combine
data flow information arising out of the same program point, resulting
in bounded representations of heap data. These representations are
simple, precise, and natural. 

The dynamically varying mapping between access expressions and
l-values is handled by abstracting out regions in the heap which can
possibly be accessed by a program.  These regions are represented by
sets of access paths and access graphs which are manipulated using a
carefully chosen set of operations.  The computation of access graphs
and access paths using data flow analysis is possible because of their
finiteness and the monotonicity of the chosen operations.  We define
data flow analyses for liveness, availability and
anticipability of heap references.  Liveness analysis is an
any path problem, hence it involve unbounded information requiring
access graphs as data flow values. Availability and anticipability
analyses are all paths problems, hence they involve bounded
information which is represented by finite sets of access paths.

\item {\em Identifying the minimal information which covers every live link
  in the heap}. An interesting aspect of our liveness analysis is that the
  property of explicit liveness captures the minimal information which covers
  every link which can possibly be live. Complete liveness is computed by
  incorporating alias information in explicit liveness.
\end{itemize}

An immediate application of these analyses is a technique to improve
garbage collection. This technique works by identifying objects which
are dead and rendering them unreachable by setting them to null as
early as possible. Though this idea was previously {known} to yield
benefits~{\cite{gcfaq}}, nullification of dead objects was based on
profiling~{\cite{shah01,shah02}}.  Our method, instead, is based on
static analysis.

For the future work, we find some scope of improvements on both conceptual 
level and at the level of implementation.
\begin{enumerate}
\item {\em Conceptual Aspects\/}.
\begin{enumerate}
\item Since the scalability of  our method critically depends on the scalability
      of alias  analysis, we would like  to explore the  possibility of avoiding
      computation  of complete alias  information at  each program  point. Since
      explicit  liveness  does not  require  alias  information, an  interesting
      question for further investigation is:  Just how much alias information is
      enough to compute complete liveness from explicit liveness?  This question
      is important because:
      \begin{itemize} 
      \item Not all aliases contribute to complete liveness.
      \item  Even  when  an  alias  contributes  to liveness,  it  needs  to  be
            propagated over a limited region of the program.
      \end{itemize}
      
      \item  We have  proposed an  efficient version  of  call strings
      based  interprocedural  data  flow  analysis in  an  independent
      work~\cite{bageshri07phd}.    It  is   a  generic
      approach which  retains full context sensitivity.  We would like
      to use it for heap reference analysis.
      \item We would like to improve the \NULL\ insertion algorithm so
        that the same link is not nullified more than once.
      \item  We  would like  to  analyze  array  fragments instead  of
        treating an entire array as  a scalar (and hence, all elements
        as equivalent).
      \item  We  would also  like to  extend the  scope  of heap
	reference analysis for functional languages.  The basic method
	and  the   details  of  the  liveness   analysis  are  already
	finalized~\cite{karkare07liveness}.   The   details  of  other
	analyses are being finalized~\cite{karkare07hra}.

\end{enumerate}
\item {\em Implementation Related Aspects\/}.
  \begin{enumerate}
  \item We would also like  to implement this approach for C/C++
    and use it for plugging memory leaks statically.
  \item  Our  experience  with  our  proof-of-concept  implementations
    indicates that the engineering  choices made in the implementation
    have a significant  bearing on the performance of  our method. For
    example, we would like to use a better representation than the one
    provided by SOOT.
  \end{enumerate}
\end{enumerate}

We would also like to apply the summarization heuristic to other
analyses. Our initial explorations indicate that a similar approach
would be useful for extending static inferencing of flow-sensitive
monomorphic types~\cite{kdm.typeinferencing} to include polymorphic
types.  This is possible because polymorphic types represent an
infinite set of types and hence discovering them requires
summarizing unbounded information.

\begin{acks}
Several people have contributed to this work in past few years. We would
particularly like to thank Smita  Ranjan, Asit Varma, Deepti Gupta, Neha
Adkar,  C.\ Arunkumar,  Mugdha Jain,  and  Reena Kharat.  Neha's work  was
supported  by Tata  Infotech Ltd.   A initial  version of  the prototype
implementation of  this work was  partially supported by  Synopsys India
Ltd.  Amey Karkare  has been  supported  by Infosys  Fellowship. We  are
thankful to the anonymous referee for detailed remarks.
\end{acks}

\bibliography{gc}

\appendix

\section{Non-Distributivity in Heap Reference Analysis}

\label{sec:non-distributivity} 

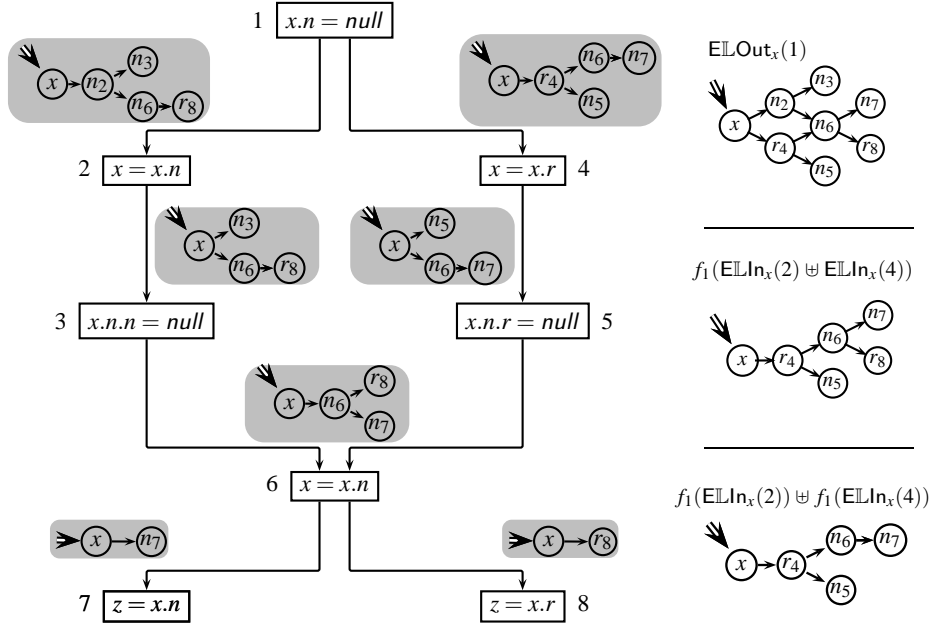
\begin{figure}[t]
\newcommand{\glx}{
\scalebox{1}{%
	\raisebox{1.4mm}{\rnode{n0}{}} \ \ \ 
	\raisebox{.6mm}{\circlenode[framesep=.8mm]{n1}{$x$}}
	\ncline[doubleline=true]{->}{n0}{n1}
	\  \  \ \ 
	\raisebox{.8mm}{\circlenode[framesep=0]{n2}{$n_7$}}
	\ncline{->}{n1}{n2}%
	\rule{0cm}{.4cm}%
}}
\newcommand{\grx}{
\scalebox{1}{%
	\raisebox{1.4mm}{\rnode{n0}{}} \ \ \ 
	\raisebox{.6mm}{\circlenode[framesep=.8mm]{n1}{$x$}}
	\ncline[doubleline=true]{->}{n0}{n1}
	\  \  \ \ 
	\raisebox{.8mm}{\circlenode[framesep=0]{n2}{$r_8$}}
	\ncline{->}{n1}{n2}%
	\rule{0cm}{.4cm}%
}}
\newcommand{\gnnrx}{
\scalebox{1}{%
\psset{unit=1mm}
\begin{pspicture}(-3,1)(17,11)
\psrelpoint{origin}{n00}{-2}{11}
\rput(\x{n00},\y{n00}){\rnode{n00}{}}
\psrelpoint{n00}{n01}{4}{-5}
\rput(\x{n01},\y{n01}){\rnode{n01}{\pscirclebox[framesep=.8]{$x$}}}
\psrelpoint{n01}{n0}{6}{0}
\rput(\x{n0},\y{n0}){\rnode{n0}{\pscirclebox[framesep=0]{$n_6$}}}
\psrelpoint{n0}{n1}{6}{3}
\rput(\x{n1},\y{n1}){\rnode{n1}{\pscirclebox[framesep=.2]{$r_8$}}}
\ncline{->}{n0}{n1}
\psrelpoint{n0}{n1}{6}{-3}
\rput(\x{n1},\y{n1}){\rnode{n1}{\pscirclebox[framesep=0]{$n_7$}}}
\ncline{->}{n0}{n1}
\ncline[doubleline=true]{->}{n00}{n01}
\ncline{->}{n01}{n0}
\end{pspicture}
}}

\newcommand{\gnrx}{
\scalebox{1}{%
\psset{unit=1mm}
\begin{pspicture}(-3,1)(17,11)
\psrelpoint{origin}{n00}{-2}{11}
\rput(\x{n00},\y{n00}){\rnode{n00}{}}
\psrelpoint{n00}{n01}{4}{-5}
\rput(\x{n01},\y{n01}){\rnode{n01}{\pscirclebox[framesep=.8]{$x$}}}
\psrelpoint{n01}{n0}{6}{-3}
\rput(\x{n0},\y{n0}){\rnode{n0}{\pscirclebox[framesep=0]{$n_6$}}}
\psrelpoint{n0}{n1}{6}{0}
\rput(\x{n1},\y{n1}){\rnode{n1}{\pscirclebox[framesep=.2]{$r_8$}}}
\ncline{->}{n0}{n1}
\psrelpoint{n01}{n2}{6}{3}
\rput(\x{n2},\y{n2}){\rnode{n2}{\pscirclebox[framesep=0]{$n_3$}}}
\ncline{->}{n0}{n1}
\ncline[doubleline=true]{->}{n00}{n01}
\ncline{->}{n01}{n0}
\ncline{->}{n01}{n2}
\end{pspicture}
}}

\newcommand{\gnnx}{
\scalebox{1}{%
\psset{unit=1mm}
\begin{pspicture}(-3,1)(17,11)
\psrelpoint{origin}{n00}{-2}{11}
\rput(\x{n00},\y{n00}){\rnode{n00}{}}
\psrelpoint{n00}{n01}{4}{-5}
\rput(\x{n01},\y{n01}){\rnode{n01}{\pscirclebox[framesep=.8]{$x$}}}
\psrelpoint{n01}{n0}{6}{-3}
\rput(\x{n0},\y{n0}){\rnode{n0}{\pscirclebox[framesep=0]{$n_6$}}}
\psrelpoint{n0}{n1}{6}{0}
\rput(\x{n1},\y{n1}){\rnode{n1}{\pscirclebox[framesep=.2]{$n_7$}}}
\ncline{->}{n0}{n1}
\psrelpoint{n01}{n2}{6}{3}
\rput(\x{n2},\y{n2}){\rnode{n2}{\pscirclebox[framesep=0]{$n_5$}}}
\ncline{->}{n0}{n1}
\ncline[doubleline=true]{->}{n00}{n01}
\ncline{->}{n01}{n0}
\ncline{->}{n01}{n2}
\end{pspicture}
}}

\newcommand{\gNnrx}{
\scalebox{1}{%
\psset{unit=1mm}
\begin{pspicture}(-3,1)(22,12)
\psrelpoint{origin}{n00}{-2}{11}
\rput(\x{n00},\y{n00}){\rnode{n00}{}}
\psrelpoint{n00}{n000}{4}{-5}
\rput(\x{n000},\y{n000}){\rnode{n000}{\pscirclebox[framesep=.8]{$x$}}}
\psrelpoint{n000}{n01}{6}{0}
\rput(\x{n01},\y{n01}){\rnode{n01}{\pscirclebox[framesep=.1]{$n_2$}}}
\psrelpoint{n01}{n0}{6}{-3}
\rput(\x{n0},\y{n0}){\rnode{n0}{\pscirclebox[framesep=0]{$n_6$}}}
\psrelpoint{n0}{n1}{6}{0}
\rput(\x{n1},\y{n1}){\rnode{n1}{\pscirclebox[framesep=.2]{$r_8$}}}
\ncline{->}{n0}{n1}
\psrelpoint{n01}{n2}{6}{3}
\rput(\x{n2},\y{n2}){\rnode{n2}{\pscirclebox[framesep=0]{$n_3$}}}
\ncline{->}{n0}{n1}
\ncline[doubleline=true]{->}{n00}{n000}
\ncline{->}{n01}{n0}
\ncline{->}{n01}{n2}
\ncline{->}{n000}{n01}
\end{pspicture}
}}

\newcommand{\gRnnx}{
\scalebox{1}{%
\psset{unit=1mm}
\begin{pspicture}(-3,-1)(22,11)
\psrelpoint{origin}{n00}{-2}{10}
\rput(\x{n00},\y{n00}){\rnode{n00}{}}
\psrelpoint{n00}{n000}{4}{-5}
\rput(\x{n000},\y{n000}){\rnode{n000}{\pscirclebox[framesep=.8]{$x$}}}
\psrelpoint{n000}{n01}{6}{0}
\rput(\x{n01},\y{n01}){\rnode{n01}{\pscirclebox[framesep=.1]{$r_4$}}}
\psrelpoint{n01}{n0}{6}{3}
\rput(\x{n0},\y{n0}){\rnode{n0}{\pscirclebox[framesep=0]{$n_6$}}}
\psrelpoint{n0}{n1}{6}{0}
\rput(\x{n1},\y{n1}){\rnode{n1}{\pscirclebox[framesep=.2]{$n_7$}}}
\ncline{->}{n0}{n1}
\psrelpoint{n01}{n2}{6}{-3}
\rput(\x{n2},\y{n2}){\rnode{n2}{\pscirclebox[framesep=0]{$n_5$}}}
\ncline{->}{n0}{n1}
\ncline[doubleline=true]{->}{n00}{n000}
\ncline{->}{n01}{n0}
\ncline{->}{n01}{n2}
\ncline{->}{n000}{n01}
\end{pspicture}
}}
\newcommand{\lout}{
\scalebox{1}{%
\psset{unit=1mm}
\begin{pspicture}(-3,-4)(22,18)
\psrelpoint{origin}{n0}{-2}{11}
\rput(\x{n0},\y{n0}){\rnode{n0}{}}
\psrelpoint{n0}{n1}{3}{-6}
\rput(\x{n1},\y{n1}){\rnode{n1}{\pscirclebox[framesep=1]{$x$}}}
\psrelpoint{n1}{n2}{6}{3}
\rput(\x{n2},\y{n2}){\rnode{n2}{\pscirclebox[framesep=.2]{$n_2$}}}
\psrelpoint{n2}{n3}{6}{3}
\rput(\x{n3},\y{n3}){\rnode{n3}{\pscirclebox[framesep=.2]{$n_3$}}}
\psrelpoint{n1}{r4}{6}{-3}
\rput(\x{r4},\y{r4}){\rnode{r4}{\pscirclebox[framesep=.4]{$r_4$}}}
\psrelpoint{r4}{n5}{6}{-3}
\rput(\x{n5},\y{n5}){\rnode{n5}{\pscirclebox[framesep=.2]{$n_5$}}}
\psrelpoint{r4}{n6}{6}{3}
\rput(\x{n6},\y{n6}){\rnode{n6}{\pscirclebox[framesep=.2]{$n_6$}}}
\psrelpoint{n6}{n7}{6}{3}
\rput(\x{n7},\y{n7}){\rnode{n7}{\pscirclebox[framesep=.2]{$n_7$}}}
\psrelpoint{n6}{r8}{6}{-3}
\rput(\x{r8},\y{r8}){\rnode{r8}{\pscirclebox[framesep=.2]{$r_8$}}}
\ncline[doubleline=true]{->}{n0}{n1}
\ncline[nodesep=-.5]{->}{n1}{n2}
\ncline[nodesep=-.5]{->}{n1}{r4}
\ncline[nodesep=-.5]{->}{n2}{n3}
\ncline[nodesep=-.5]{->}{n2}{n6}
\ncline[nodesep=-.5]{->}{r4}{n5}
\ncline[nodesep=-.5]{->}{r4}{n6}
\ncline[nodesep=-.5]{->}{n6}{n7}
\ncline[nodesep=-.5]{->}{n6}{r8}
\psrelpoint{origin}{n0}{4}{15}
\rput(\x{n0},\y{n0}){\rnode{n0}{$\Lout{x}(1)$}}
\end{pspicture}
}}

\newcommand{\lina}{
\scalebox{1}{%
\psset{unit=1mm}
\begin{pspicture}(-3,-3)(22,18)
\psrelpoint{origin}{n0}{-2}{10}
\rput(\x{n0},\y{n0}){\rnode{n0}{}}
\psrelpoint{n0}{n1}{4}{-6}
\rput(\x{n1},\y{n1}){\rnode{n1}{\pscirclebox[framesep=1]{$x$}}}
\psrelpoint{n1}{r4}{6}{0}
\rput(\x{r4},\y{r4}){\rnode{r4}{\pscirclebox[framesep=.4]{$r_4$}}}
\psrelpoint{r4}{n5}{6}{-3}
\rput(\x{n5},\y{n5}){\rnode{n5}{\pscirclebox[framesep=.2]{$n_5$}}}
\psrelpoint{r4}{n6}{6}{3}
\rput(\x{n6},\y{n6}){\rnode{n6}{\pscirclebox[framesep=.2]{$n_6$}}}
\psrelpoint{n6}{n7}{6}{3}
\rput(\x{n7},\y{n7}){\rnode{n7}{\pscirclebox[framesep=.2]{$n_7$}}}
\psrelpoint{n6}{r8}{6}{-3}
\rput(\x{r8},\y{r8}){\rnode{r8}{\pscirclebox[framesep=.2]{$r_8$}}}
\ncline[doubleline=true]{->}{n0}{n1}
\ncline[nodesep=-.5]{->}{n1}{r4}
\ncline[nodesep=-.5]{->}{r4}{n5}
\ncline[nodesep=-.5]{->}{r4}{n6}
\ncline[nodesep=-.5]{->}{n6}{n7}
\ncline[nodesep=-.5]{->}{n6}{r8}
\psrelpoint{origin}{n0}{10}{16}
\rput(\x{n0},\y{n0}){\rnode{n0}{$f_1 (\Lin{x}(2) \cupG \Lin{x}(4))$}}
\end{pspicture}
}}

\newcommand{\linb}{
\scalebox{1}{%
\psset{unit=1mm}
\begin{pspicture}(-3,4)(24,20)
\psrelpoint{origin}{n0}{-20}{0}
\rput(\x{n0},\y{n0}){{\scalebox{1.1}{\gRnnx}\ \ \ \ \ \ \ \ \ \ \ }}
\psrelpoint{origin}{n0}{10}{17}
\rput(\x{n0},\y{n0}){\rnode{n0}{$f_1 (\Lin{x}(2)) \cupG f_1(\Lin{x}(4))$}}
\end{pspicture}
}}

\psset{unit=1mm}
\begin{tabular}{cc}
\begin{pspicture}(0,15)(90,110)
\small
\psrelpoint{origin}{n1}{45}{95}
\rput(\x{n1},\y{n1}){\rnode{n1}{1\ \ \psframebox{$x.n=\NULL$}\white \ \ 1}}
\psrelpoint{n1}{n2}{-25}{-20}
\rput(\x{n2},\y{n2}){\rnode{n2}{2\ \ \psframebox{$x=x.n$}\white \ \ 2}}
\psrelpoint{n1}{n3}{25}{-20}
\rput(\x{n3},\y{n3}){\rnode{n3}{\white 4\ \ \black\psframebox{$x=x.r$}\ \ 4}}
\psrelpoint{n2}{n4}{0}{-20}
\rput(\x{n4},\y{n4}){\rnode{n4}{3\ \ \psframebox{$x.n.n=\NULL$}\white \ \ 3}}
\psrelpoint{n3}{n5}{0}{-20}
\rput(\x{n5},\y{n5}){\rnode{n5}{\white 5\ \ \black\psframebox{$x.n.r=\NULL$}\ \ 5}}
\psrelpoint{n5}{n6}{-25}{-22}
\rput(\x{n6},\y{n6}){\rnode{n6}{6\ \ \psframebox{$x=x.n$}\white \ \ 6}}
\psrelpoint{n6}{n7}{-25}{-16}
\rput(\x{n7},\y{n7}){\rnode{n7}{7\ \ \psframebox{$z=x.n$}\white \ \ 7}}
\psrelpoint{n7}{n71}{-10}{3}
\rput(\x{n7},\y{n7}){\rnode{n7}{7\ \ \psframebox{$z=x.n$}\white \ \ 7}}
\psrelpoint{n6}{n8}{25}{-16}
\rput(\x{n8},\y{n8}){\rnode{n8}{\white 8\ \ \black\psframebox{$z=x.r$}\ \ 8}}
\ncdiag[armA=12.25,armB=3.3,linearc=.25,offsetA=-2,angleA=270,angleB=90]{->}{n1}{n2}
\ncdiag[armA=12.25,armB=3.3,linearc=.25,offsetA=2,angleA=270,angleB=90]{->}{n1}{n3}
\ncline{->}{n2}{n4}
\ncdiag[armA=14.4,armB=3.1,linearc=.25,offsetB=-2,angleA=270,angleB=90]{->}{n4}{n6}
\ncdiag[armA=14.4,armB=3.1,linearc=.25,offsetB=2,angleA=270,angleB=90]{->}{n5}{n6}
\ncline{->}{n3}{n5}
\ncdiag[armA=9.2,armB=2.6,linearc=.25,offsetA=-2,angleA=270,angleB=90]{->}{n6}{n7}
\ncdiag[armA=9.2,armB=2.6,linearc=.25,offsetA=2,angleA=270,angleB=90]{->}{n6}{n8}
\psrelpoint{n7}{np}{-5}{9}
\rput(\x{np},\y{np}){\psframebox[framesep=0,linestyle=none,framearc=.5,fillstyle=solid,fillcolor=lightgray]{\glx}}
\psrelpoint{n8}{np}{5}{9}
\rput(\x{np},\y{np}){\psframebox[framesep=0,linestyle=none,framearc=.5,fillstyle=solid,fillcolor=lightgray]{\grx}}
\psrelpoint{n6}{np}{-1}{11}
\rput(\x{np},\y{np}){\psframebox[framesep=0,linestyle=none,framearc=.5,fillstyle=solid,fillcolor=lightgray]{\gnnrx}}
\psrelpoint{n4}{np}{12}{10}
\rput(\x{np},\y{np}){\psframebox[framesep=0,linestyle=none,framearc=.5,fillstyle=solid,fillcolor=lightgray]{\gnrx}}
\psrelpoint{n5}{np}{-12}{10}
\rput(\x{np},\y{np}){\psframebox[framesep=0,linestyle=none,framearc=.5,fillstyle=solid,fillcolor=lightgray]{\gnnx}}
\psrelpoint{n3}{np}{-55}{12}
\rput(\x{np},\y{np}){\psframebox[framesep=0,linestyle=none,framearc=.5,fillstyle=solid,fillcolor=lightgray]{\gNnrx}}
\psrelpoint{n3}{np}{5}{12}
\rput(\x{np},\y{np}){\psframebox[framesep=0,linestyle=none,framearc=.5,fillstyle=solid,fillcolor=lightgray]{\gRnnx}}
\end{pspicture}
&\begin{tabular}[b]{@{}l@{}}
\lout\\ \\ \hline\\
\lina\\ \\ \hline\\
\linb \\\\
\end{tabular}
\end{tabular}
\caption{Non-distributivity of liveness analysis.  Access path
\mbox{$x\protect\myarrow r\protect\myarrow n\protect\myarrow r$} is a
spurious access path which does not get killed by the assignment in
block 1.}
\label{fig:liveness.non-distributivity}
\rule{\textwidth}{.2mm}
\end{figure}

Explicit liveness analysis defined in this paper is not distributive
whereas availability and anticipability analyses are distributive.
%
%
Explicit liveness analysis
is non-distributive because of the approximation introduced by the $\cupG$
operation.  \mbox{$G_1 \cupG G_2$} may contain access paths which are
neither in $G_1$ nor in $G_2$.
\begin{example}
\label{exmp:non-distributivity.liveness}
Figure~\ref{fig:liveness.non-distributivity} illustrates the
non-distributivity of explicit liveness analysis.  Liveness graphs associated
with the entry each block is shown in shaded boxes.  Let $f_1$ denote
the flow function which computes $x$-rooted liveness graphs at the
entry of block 1.  Neither $\Lin{x}(2)$ nor $\Lin{x}(4)$ contains the
access path \mbox{$x\myarrow r\myarrow n\myarrow r$} but their union
contains it.  It is easy to see that
\[
 f_1 (\Lin{x}(2) \cupG \Lin{x}(4)) \sqsubseteq_G
  f_1 (\Lin{x}(2))  \cupG f_1(\Lin{x}(4)) \\
\]
\mybox
\end{example}

 Availability  and  anticipability analyses  are  non-distributive because  they
depend on may-alias analysis which is non-distributive.

\end{document}